\documentclass[11pt]{article}
\usepackage{epsfig}
\usepackage{latexsym}
\usepackage{amsmath}
\usepackage{amssymb}
\usepackage{amsthm}
\usepackage{enumerate}
\usepackage[all]{xy}

\newtheorem{theorem}{Theorem}{\bf}{\it}
\newtheorem*{prop_equiv_rel}{Proposition \ref{prop:sim_equiv}}{\bf}{\it}
\newtheorem*{prop_sim_fiber}{Proposition \ref{prop:sim_fiber}}{\bf}{\it}
\newtheorem*{prop_GL_i}{Theorem \ref{thm:G_L}.(i)}{\bf}{\it}
\newtheorem*{prop_GL_ii}{Theorem \ref{thm:G_L}.(ii)}{\bf}{\it}
\newtheorem*{prop_GL_iii}{Theorem \ref{thm:G_L}.(iii)}{\bf}{\it}
\newtheorem*{lem_adjoint}{Lemma \ref{lem:adjoint_action}}{\bf}{\it}
{\bf}{\rm}
\newtheorem{proposition}[theorem]{Proposition}{\bf}{\it}
\newtheorem{corollary}[theorem]{Corollary}{\bf}{\it}
\newtheorem{lemma}[theorem]{Lemma}{\bf}{\it}

\renewcommand{\forall}{\mbox{for all}\,\,}

\def\bb{{\bf b}} 
\def\be{{\bf e}}
\def\C{{\mathfrak C}} 
\def\mc{{\mathfrak c}}\def\md{{\mathfrak d}}
\def\D{{\cal D}} \def\G{{\bf G}} \def\H{{\cal H}}
\def\eps{\varepsilon} \def\M{{\bf M}}\def\N{N}
\def\bO{{\bf O}}
  \def\fphi{\Phi}
\def\phi{\varphi}

\def\munt{{\underline{m}}}\def\nunt{{\underline{n}}}
\def\runt{\underline{r}}\def\circe{{e_0}}
\def\stab{{\mathfrak S}}

\def\betavon{\hat\beta}
\def\rhovon{\hat\rho}
\def\trho{\tilde\rho}
\def\tbeta{\tilde\beta}

\def\tlambda{\tilde\lambda}
\def\ttau{\tilde\tau}
\def\rot{\rho}
\def\rotb{\sigma}
\def\boost{\beta}
\def\boostb{\gamma}
\def\hatpi{\bar\pi}
\def\hatsim{\bar\sim}
\def\hatz{\bar Z}
\def\hate{\bar e}
\def\hatf{\bar f}
\def\hatzeta{\bar\zeta}
\def\upparrow{+}

  \def\Aut{{\rm Aut}}
\def\ba{{\bf a}} \def\bF{{\bf F}} \def\complex{{\mathbb C}}
 \def\epsilon{\varepsilon}
\def\half{\mbox{$\frac{1}{2}$}}

  \def\supp{{\rm supp}}
 
\def\test{{\mathfrak D}}

\def\integers{{\mathbb Z}} \def\complex{{\mathbb C}}
\def\reals{{\mathbb R}}
\def\({\left(}
\def\){\right)}

\title{Spin, Statistics, and Reflections\\II. Lorentz
  Invariance\thanks{Dedicated to Professor H.-J. Borchers on the
    occasion of his 80th birthday}} 

\author{Bernd Kuckert\thanks{Emmy-Noether fellow of the Deutsche
    Forschungsgemeinschaft}\\ Reinhard Lorenzen\\
  II. Institut f\"ur Theoretische Physik\\ Luruper Chaussee 149, 22761
  Hamburg, Germany}

\begin{document}

\maketitle
\begin{abstract}
  The analysis of the relation between modular P$_1$CT-symmetry --- a
  consequence of the Unruh effect --- and Pauli's spin-statistics
  relation is continued. The result in the predecessor to this article
  is extended to the Lorentz symmetric situation. A model $\G_L$ of
  the universal covering $\widetilde{L_+^\uparrow}\cong
  SL(2,\complex)$ of the restricted Lorentz group $L_+^\uparrow$ is
  modelled as a reflection group at the classical level. Based on this
  picture, a representation of $\G_L$ is constructed from pairs
  of modular P$_1$CT-conjugations, and this representation can easily
  be verified to satisfy the spin-statistics relation.
\end{abstract}

\section{Introduction}

The spin-statistics connection and the search for its conceptual roots
has been a prominent object of investigation in quantum field theory
over decades; we refer to Refs. \ref{StW}, \ref{Kuc04a}, and \ref{Kuc04}
for detailed discussions of the literature in this field. Spectacular
success has been made in deriving the spin-statistics connection from
standard properties of quantum fields, but there has always remained
some dissatisfaction because these results did not really dig up the
physical roots of the principle. Recently, an angular-momentum
additivity condition has been established as sufficient and necessary
for Pauli's spin-statistics connection in quantum mechanics
\cite{Kuc04a,KM}, but this result does not include quantum fields,
which will be of interest here.

In particular, the analysis was confined to finite-component fields,
which entails a strong assumption on the representation of the Lorentz
group one wishes to investigate. There are, however,
infinite-component quantum fields that are covariant under
representations {\it violating} Pauli's spin-statistics connection
\cite{Str}. What is more, the confinement to finite-component fields
is of a purely technical nature; there is no evident physical
motivation except the fact that it is met in practically all
applications. So despite the merits of the old results, one agreed
that some work remained to be done.

In the 1990s, it was realized by several authors
\cite{FM,GL95,Kuc95,GL96} that the spin-statistics connection could be
derived from the Unruh effect \cite{Unr,BW75,BW76}. This phenomenon,
which states that a uniformly accelerated observer experiences the
vacuum of a quantum field as a thermal state, has recently been
derived from basic stability properties of vacuum states \cite{Kuc02}.

The Unruh effect, in turn, implies an intrinsic form of
P$_1$CT-symmetry, i.e., covariance under conjugations in charge, time,
and {\it one} spatial direction \cite{GL95}. These conjugations can be
extracted from the algebra of field operators by an elementary
intrinsic fashion invented by Tomita and Takesaki \cite{Tak70,BR1}.
This symmetry is referred to as {\it modular P$_1$CT-symmetry}.

In 1994/95, two spin-statistics theorems were obtained by Guido and
Longo on the one hand \cite{GL95} and by one of us on the other
\cite{Kuc95}. Guido and Longo derived the spin-statistics theorem
from the Unruh effect, and their result applies to a large class of
quantum field theories, including fields with an infinite number of
components and massless fields. On the other hand, the result found in
Ref \ref{Kuc95} merely assumes modular P$_1$CT-symmetry in the vacuum
sector of a Haag-Kastler theory of observables, and it deduced the
spin-statistics connection for massive single-particle states, which
give rise to topological charges. The elements of the representation
of the (homogeneous) symmetry group are products of two modular
conjugations each, which yields an elementary algebraic argument. But
massless fields are not included, and the setting prevents the
observables to be covariant under more than one representation of the
Poincar\'e group. So one result covers a larger class of fields by
making stronger symmetry assumptions, whereas the other one minimizes
the symmetry assumptions by considering a smaller class of fields.

The result presented in this article joins the advantages of these two
approaches: Only modular P$_1$CT-symmetry is assumed, and the result
covers fields satisfying an absolute minimum of standard assumptions.
Not even covariance under a representation of the Lorentz group needs
to be assumed from the outset; this representation will be {\it
  constructed} from the modular P$_1$CT-operators.

As an important prerequisite, a model of the universal covering group
of the restricted Lorentz group will be constructed first. The model
is a kind of a reflection group.

In Ref. \ref{Kuc04}, a model $\G_R$ of the universal covering group
$\widetilde{SO(3)}\cong SU(2)$ has been constructed from pairs of
reflections at planes in $\reals^3$. Considering a general quantum
field theory, it was assumed that modular P$_1$CT-conjugations existed
for all reflections along spacelike vectors in a fixed time-zero
plane. This symmetry assumption has been shown to be sufficient to
construct a covariant representation of $\G_R$, and it is elementary
to see that this representation exhibits the spin-statistics relation.

For the restricted Lorentz group $L_+^\uparrow$, such a representation has been
constructed earlier by Buchholz, Dreyer, Florig, and Summers
\cite{BDFS,Flo}. In a more recent article, Buchholz and Summers have given
a much more straightforward proof \cite{BS}. The short cut found there
was the decisive indication that a similar result could also be
obtained for the universal covering $\widetilde L_+^\uparrow\cong
SL(2,\complex)$, the goal being the generalization of the first
derivation of the spin-statistics theorem already obtained in Ref.
\cite{Kuc95}. Some of their arguments will play an important role at
the end of this paper, where such a generalization is established.

This article will be subdivided as follows. In Section
\ref{sect:prel}, some preliminaries will be discussed, in Section
\ref{subsec:G_L_statements}, the construction of the covering group $\G_L$ will be
outlined in terms of definitions and statements, which will be proved
in Section \ref{sect:proofs}. The construction of $\G_L$ will be
applied when proving a most general spin-statistics theorem for
relativistic quantum fields, which is done in Section \ref{sect:spst}.
In Section \ref{sect:other}, it is shown that modular P$_1$CT-symmetry
implies full PCT-symmetry as well and how the present result is
related to the results of Guido and Longo obtained in Ref.
\ref{GL95}. The article ends with a conclusion.

\subsection{Preliminaries}\label{sect:prel}

Let $\reals^{1+3}$ be the Minkowski spacetime with three spatial
dimensions, denote by $g(\cdot,\cdot)$ its Lorentz metric $(x,y)\mapsto
g(x,y)=:xy$, by $V_+$ the open forward light cone,\footnote{The set 
$\{x\in\reals^{1+s}:\,x^2>0\}$ has two connected components. The open 
forward lightcone $V_+$ is the future-directed one with respect to the 
time orientation of $\reals^{1+3}$.}
by $M_1^+$ the hyperboloid $\{x\in V_+:\,x^2=1\}$, and by $H_1$ the
spacelike unit hyperboloid $\{x\in\reals^{1+3}:\,x^2=-1\}$.

The Lorentz group $L$ has four connected components. The connected
component $L_+^\uparrow=:L_1$ containing the unit element $1$ is a 
subgroup of $L$ called the {\it restricted Lorentz group}. All 
$\mu\in L_1$ satisfy $\det\mu=1$ and
$\mu V_+=V_+$. The fixed-point set $FP(\mu)$ of any $\mu\in L_1$ is a
linear subspace of $\reals^{1+3}$ with zero, one, two, or four
dimensions.

We call $\mu\in L_1$ a {\it generalized boost} if $FP(\mu)$ contains a
two-dimensional spacelike subspace, and we call $\mu$ a {\it
  generalized rotation} if $FP(\mu)$ contains a two-dimensional
timelike subspace. The usual notions of boost and rotation require the
choice of a time vector $e_0\in M_1^+$ and its time-zero plane
$e_0^\perp$.  A generalized boost $\mu$ is a boost with respect to
$e_0$ if $FP(\mu)\subset e_0^\perp$, and it is a rotation if
$FP(\mu)^\perp \subset e_0^\perp$. For each generalized rotation or
boost $\mu$ there is more than one $e_0\in M_1^+$ with respect to
which $\mu$ is a rotation or boost, respectively.

Note that the unit element is both a generalized boost and a
generalized rotation and that the fixed-point sets of all other
generalized rotations and boosts are two-dimensional.

Crucial for the analysis to follow is the fact that each element of
$L_1$ is a concatenation of two orthogonal reflections at
two-dimensional spacelike planes.\footnote{see Lemma
  \ref{lem:lambda_onto} below.} Like in Ref. \ref{Kuc04}, where the
corresponding analysis was carried out for the simpler case of
rotational symmetry, a simply connected covering of $L_1$ will now be
constructed by endowing these planes with an orientation.

There are several equivalent and useful descriptions of the set $\bO$
of oriented spacelike planes.

\smallskip\smallskip
\begin{enumerate}
\item{\it Rindler wedges.} The spacelike complement
  $S'=\{x\in\reals:\,xs<0\,\forall\,s\in S\}$ of a spacelike plane $S$
  has two connected components, each of which specifies an orientation
  on $S$.  These components are wedges and have been named after
  W. Rindler, who endowed them with a spacetime structure on their own.
  The geodesic observers in this spacetime structure are those
  observers in $\reals^{1+3}$ that are uniformly accelerated
  perpendicular to $S$.  The boundary of a Rindler wedge is a horizon
  for the Rindler observer.  This physical role will be relevant in
  the discussion of the spin-statistics relation below.

\item{\it Classes of zweibeine.} \label{rem:zweibeine} Define a set of
  zweibeine $Z$ by
$$Z:=\{\xi=(t_\xi,x_\xi):\,t_\xi^2=1,x_\xi^2=-1,\,x_\xi \perp t_\xi\}.$$
The set $\xi^\perp:=\{t_\xi,x_\xi\}^\perp$ is a two-dimensional
spacelike plane, and the wedge $W_\xi:=\{x\in\reals
^{1+3}:\,xx_\xi>|xt_\xi|\}$ is one of its Rindler wedges.
Define an equivalence relation $\xi\hatsim\eta$ on $Z$ by the
condition $W_\xi=W_\eta$.

Let $\hatz$ be the quotient space $Z/\hatsim$, and let $\hatpi$ be
the canonical projection from $Z$ onto $\hatz$. For each 
$a=\hatpi(\xi)$, denote by $W_a$ the wedge $W_\xi$ and by 
$a^\perp$ its edge $\xi^\perp$.

Given $a\in\hatz$, the hyperbola
$\Gamma(a):=\{x_\xi:\,\xi\in\bar\pi^{-1}(a)\}$ is a geodesic of the
Rindler spacetime structure on $W_a$.

An action of the full Lorentz group $L$ on $Z$ from the left is
defined by $\mu\xi=(\mu t_\xi,\mu x_\xi)$. Since $\xi\hatsim\eta$
implies $\mu\xi\hatsim\mu\eta$, this action induces an action on
$\hatz$ by $\mu\hatpi(\xi):=\hatpi(\mu\xi)$. Evidently, $W_{\mu a}=\mu
W_a$.

The subset $Z^{+}:=\{\xi\in Z:\,t_\xi\in V_+\}$ of $Z$ has the
property that $\hatpi(Z^{+})=\hatpi(Z)$. If one restricts the
equivalence relation $\hatsim$ to $Z^{+}$, one obtains an
equivalence relation as well, and the corresponding quotient space is
isomorphic with $\hatz$. The restricted Lorentz group $L_1$ acts
transitively on $Z^{+}$ (see Lemma \ref{lem:transitive} below),
but not on $Z$ (since the elements of $L_1$ preserve time
orientation). 

\item{\it Spectral decompositions of boosts.} \label{item:spec-dec}
  Given $a\in\hatz$, the generalized boosts with fixed-point set
  $a^\perp$ give rise to a one-parameter group $(\mu^a_\chi)_\chi$
  with the property that $(\mu_\chi^a x-x)^2>0$ for all $\chi>0$ and
  $x\in W_a$. This group is unique up to multiplication of $\chi$ by
  a positive scalar.
  
  Conversely, given a one-parameter group $(\mu_\chi)_{\chi\in\reals}$
  of generalized boosts, the $\mu_\chi$ with $\chi\neq0$ have a common
  fixed-point plane $S$. Furthermore, one verifies that there are an
  $\alpha>0$ and a future-directed lightlike vector $\ell^{+}$ with
  $\mu_\chi\ell^{+}=e^{\alpha\chi}\ell^{+}$ for all $\chi\in\reals$,
  and a past-directed lightlike vector $\ell^{-}$ with
  $\mu_\chi\ell^{-}=e^{-\alpha\chi}\ell^{-}$ for all $\chi\in\reals$.
  The convex hull of $\ell^{{+}}$, $\ell^{-}$, and $S$ is the closure
  of a Rindler wedge.
  
  Pairs of lightlike vectors have been used earlier by Buchholz,
  Dreyer, Florig, and Summers for the description of Rindler wedges
  \cite{BDFS}.
\end{enumerate}

The subsequent analysis will be formulated in terms of $\hatz$ rather
than the naturally isomorphic set $\bO$, but occasionally, the other
descriptions show up in the proofs as well.

\subsection{The construction of $\G_L$: definitions and statements}
\label{subsec:G_L_statements}

For each $\xi\in Z$, let both $j_\xi$ and $j_{\hatpi(\xi)}$ denote the
orthogonal reflection by the plane $\xi^\perp=a^\perp$, i.e., the map
$$j_\xi x\equiv
j_{\hatpi(\xi)}x:=x-2(xt_\xi)\,t_\xi-2(xx_\xi)\,x_\xi.$$
  
Endow the set $\hatz\times\hatz=:{\M}_{L}$ with the structure of the
pair groupoid of $\hatz$, and define an operation of $L$ on $\M_L$ by
$\mu(a,b):=(\mu a,\mu b)$.  With each $(a,b)\in\M_L$, one can
associate the Lorentz transformation $\lambda(a,b):= j_{a}j_{b}\in
L_1$. Define a relation $\sim$ on $\M_L$ by writing $\munt\sim\nunt$
if and only if there exists a $\mu\in L_1$ with $\nunt=\pm\mu^2\munt$
and $\mu\lambda(\munt)\mu^{-1}=\lambda(\munt)$. Note that $\mu\sim\nu$
implies $\lambda(\mu)=\lambda(\nu)$.\,\footnote{The square in the
  condition $\nunt=\pm\mu^2\munt$ is important in order to avoid
  trouble with rotations by the angle $\pi$. It has been forgotten in
Ref. \cite{Kuc04}.}

\begin{proposition}\label{prop:sim_equiv}
The relation $\sim$ is an equivalence relation. 
\end{proposition}
The proof will be given in section \ref{sec:sim_equiv}.

Let $\G_L$ be the quotient space $\M_L/\!\!\sim$, and denote by
$\pi:\,\M_L\to\G_L$ the canonical projection of the relation $\sim$.
Define $\pm1:=\pi(a,\pm a)$ for arbitrary $a\in\hatz$, and
$-\pi(a,b):=\pi(a,-b)$ for $(a,b)\in\M_L$. 

\begin{proposition}\label{prop:sim_fiber}
  For each $g\in\G_L$, one has $g\neq-g$ and
  $\tilde\lambda^{-1}(\tilde\lambda(g))=\{g,-g\}$.
\end{proposition}
The proof of this theorem will be given in Section
\ref{sec:sim_fiber}.

As remarked, $\mu\sim\nu$ implies $\lambda(\mu)=\lambda(\nu)$, so a
map $\tilde\lambda:\,\G_L\to L_1$ can be defined by
$\tilde\lambda(g):=\lambda(\pi^{-1}(g))$, and the diagram
\begin{equation}\label{dia:lambdas}
  \xymatrix{  {\M}_{L} \ar[r]^{\pi}\ar[d]_-{\lambda} & {\G}_{L} 
\ar[dl]^-{\tilde\lambda} \\
    {L}_+^\uparrow & }
\end{equation}
commutes. All maps in this diagram are continuous. This holds for
$\pi$ by definition, and it is evident for $\lambda$. To show
continuity of $\tilde\lambda$, let $M\subset L_1$ be
open. $\tilde\lambda^{-1}(M)$ is open if and only if
$\pi^{-1}(\tilde\lambda^{-1}(M))$ is open.  This set coincides with
$\lambda^{-1}(M)$, which is open by continuity of $\lambda$.

\begin{theorem}\label{thm:G_L} \mbox{}
  \begin{enumerate}[(i)] 
  \item $\tilde\lambda$ is a covering map and endows $\G_L$ with
    the structure of a two-sheeted covering space of $L_1$.
  \item $\G_L$ is simply connected.
  \item There is a unique group product $\odot $ on ${\G}_{L}$ with
  the property that the diagram
\begin{equation}
  \xymatrix{ {\M}_{L}\times {\M}_{L}  \ar[r]^-{\circ } \ar[d]_{\pi\times\pi} 
        &  {\M}_{L}\ar[d]^{\pi} \\
  {\G}_{L}\times {\G}_{L} \ar[r]^-{\odot} 
\ar[d]_{\tilde\lambda\times\tilde\lambda} 
        & {\G}_{L} \ar[d]^{\tilde\lambda} \\
  {L}_1\times {L}_1 \ar[r]^-{\cdot}            &{L}_1 }
\end{equation}
commutes.
  \end{enumerate}
\end{theorem}
This means that $\G_L$ is isomorphic with the universal covering group
of $L_1$. The proof will be given in section \ref{sect:G_L}.

\begin{lemma}[Adjoint action of $\G_L$ on itself]\label{lem:adjoint_action}
Given $h\in\G_L$ and $(c,d)\in\M_L$, one has 
\begin{equation}
h\pi(c,d)h^{-1}=\pi\(\tlambda(h)c,\tlambda(h)d\).
\end{equation}
\end{lemma}

\section{The construction of $\G_L$:\, proofs}\label{sect:proofs}

The proofs of the statements made in the preceding section requires
an extended mathematical analysis, which will now be developed step by
step.

\subsection{Reflections of spacelike planes by spacelike planes}

It may well be that the following lemma, which is highly plausible at
a first glance, but somewhat tricky to prove, has been proved earlier
by other authors. But since such a reference is not known to us, we 
prove it here.

\begin{lemma}\label{lem:reflecting_planes}
If $A$ and $B$ are two-dimensional spacelike planes of 
$\reals^{1+3}$, then there exists a spacelike plane $C$ such that
$B$ is the image of $A$ under orthogonal reflection by $C$.
\end{lemma}

\begin{proof} If $A$ and $B$ have nontrivial intersection, then there
exist linearly independent nonzero vectors $a\in A$, $b\in B$, and
$c\in A\cap B$. The one-dimensional timelike space $\{a,b,c\}^\perp$
is perpendicular to both $A$ and $B$, so $A$ and $B$ are subspaces of
a common time-zero plane, and the problem boils down to the well-known
three-dimensional euclidean case.

It remains to consider the case that $A$ and $B$ have trivial
intersection. $A^\perp$ and $B^\perp$ are timelike planes and, hence,
are spanned by future-directed lightlike vectors $x,y\in A^\perp$ and
$v,w\in B^\perp$. Since $A$ and $B$ have trivial intersection, $x$,
$y$, $v$, and $w$ are linearly independent.

Let $C$ be the plane spanned
by the vectors $x-\alpha v$ and $y-\beta w$, where 
\[ \alpha:=\sqrt{\frac{xy}{vw}\frac{xw}{yv}}>0\qquad{\rm and}\qquad
\beta:=\sqrt{\frac{xy}{vw}\frac{yv}{xw}}>0. \]
Then $C^\perp$ is spanned by $x+\alpha v$ and $y+\beta w$, since
\[ (x-\alpha v)(x+\alpha v)=x^2-\alpha^2 v^2=0=(y-\beta w)(y+\beta w)\]
and since
\begin{align*}
  (x-\alpha v)(y&+\beta w)=xy-\alpha\beta\, vw-\alpha\,yv+\beta\,xw\\
  &=xy-
  \sqrt{\frac{xy}{vw}\frac{xw}{yv}\,\,\frac{xy}{vw}\frac{yv}{xw}}\,vw
  -\sqrt{\frac{xy}{vw}\frac{xw}{yv}}\,yv
  +\sqrt{\frac{xy}{vw}\frac{yv}{xw}}\,xw\\
  &=xy-\frac{xv}{vw}\,vw-\sqrt{\frac{xy}{vw}\,(xw)(yv)}
  +\sqrt{\frac{xy}{vw}\,(yv)(xw)}\\
  &=0=\dots=(x+\alpha v)(y-\beta w).
\end{align*}
$C^\perp$ is timelike because 
\[ (x+\alpha v)^2=x^2+2\alpha\,xv+\alpha^2v^2=2\alpha\,xv>0 \]
by $x$ and $v$ beeing future directed and by $\alpha>0$, so $C$ is spacelike.
Denote by $j_C$ the orthogonal reflection at $C$.  One then finds
\[ j_C x=\half\,j_C\left(\underbrace{(x+\alpha v)}_{\in C^\perp}+
\underbrace{(x-\alpha v)}_{\in C}\right)=\half\left(-(x+\alpha v)+(x-\alpha
v)\right)=-\alpha v\in B \]
and $j_Cy=-\beta w\in B$.
\end{proof}

\subsection{Proof of Proposition \ref{prop:sim_equiv}}\label{sec:sim_equiv}

Proposition \ref{prop:sim_equiv} states that the relation $\sim$ is an
equivalence relation. In contrast to the corresponding statement for
the analysis of the rotation group and its universal covering, this is
not self-evident. It will be proved in this section, together
with some properties of the equivalence relation $\sim$.

\begin{lemma} \label{lem:commutants}
Let $\mu$ and $\nu$ be restricted Lorentz transformations.
\begin{itemize}
\item[(i)] Suppose that $\mu \neq 1$. There exist at least one and at
  most two elements $\nu$ with $\mu = \nu^2$. If, in particular,
  $\mu^2=1$, there are two such square roots $\nu_+$ and $\nu_-$, and
  $\nu_+\nu_- = 1$.
\item[(ii)] The commutant of $\mu$ is an abelian group if and only if
  $\mu^2 \neq 1$.
\item[(iii)] Given $\mu,\nu\in L_1$, suppose that $\mu^2\neq
  1\neq\nu^2$ and $\mu^2\nu^2=\nu^2\mu^2$.  Then $\mu\nu=\nu\mu$.
\end{itemize}
\end{lemma}
\begin{proof}
  The matrix group $SL(2,\complex)$ is well known to be isomorphic
  with the universal covering group of $L_1$. Let $\Lambda$ be any
  covering map from $SL(2,\complex)$ onto $L_1$. Then
  $\Lambda^{-1}(\Lambda(A)) = \pm A$ for any $A \in SL(2,\complex)$.
  
  The conjugacy classes of $SL(2,\complex)$ are classified by the
  Jordan matrices in $SL(2,\complex)$, which are
\[N_z := \begin{pmatrix} z&0\\0&1/z\end{pmatrix}, z \in \dot\complex
\qquad N_\infty :=\begin{pmatrix} 1&1\\0&1\end{pmatrix} \qquad
N_{-\infty}:=\begin{pmatrix} -1&1\\0&-1\end{pmatrix},\] so for each $A
\in SL(2,\complex)$ there is a $z\in \dot\complex \cup \pm\infty$ and
a $P\in SL(2,\complex)$ with $A=PN_zP^{-1}$.

\smallskip\noindent {\it Proof of (i).}  Since $\mu\neq1$ by
assumption and since $[N_{-z}]=[-N_z]$, there exists an element
$A=PN_zP^{-1}\in\Lambda^{-1}(\mu)$ with $\pm1\neq z\neq-\infty$.
  
  If $z\neq\infty$, the elements of $\Lambda^{-1}(\mu)$ are $\pm A$,
  and $B_\pm\in SL(2,\complex)$ satisfy $B_\pm^2=\pm A$ if and only if
  $\pm B_\pm=\pm PN_{w_{\pm}}P^{-1}$ for complex square roots $w_\pm$
  of $\pm z$. One obtains two square roots
  $\nu_\pm:=\Lambda(B_\pm)\equiv\Lambda(-B_\pm)$ of $\mu$.
  
  If $z=\infty$, then $B:=\pm
  P\left(\begin{array}{cc}1&1/2\\0&1\end{array}\right)P^{-1}$ are the
  two square roots of $A$. Since the elements in $[N_{-\infty}]$ have
  no square roots in $SL(2,\complex)$, the only square root of $\mu$
  is $\nu:=\Lambda(B)\equiv\Lambda(-B)$.
  
  If $\mu^2=1$, then $z^2=\pm i$, so $\mu$ has two roots $\nu_+$ and
  $\nu_-$.  In order to prove $\nu_+\nu_-=1$, let $w_+$ be a square
  root of $i$, then $w_-:=\overline{w_+}$ is a square root of $-i$.
  One obtains
\[ \nu_+\nu_-=\Lambda(PN_{w_+} P^{-1})\Lambda(PN_{w_-}P^{-1})
=\Lambda(PN_{w_+} P^{-1}\,PN_{\overline{w_+}}P^{-1})=1. \]

\smallskip\noindent {\it Proof of (ii).}  $\mu\nu=\nu\mu$ if and only
if $AB=\pm BA$ for all $A\in\Lambda^{-1}(\mu)$ and
$B\in\Lambda^{-1}\nu$.

Given $A=PN_zP^{-1}\in SL(2,\complex)$ with $z\neq\pm1$, the commutant
of $A$ is the abelian group $\{PN_zP^{-1}:\,z\in\dot\complex\}$.

The anticommutant of $A$ is trivial if $z\neq\pm i$; otherwise it
consists of the matrices $P\left(\begin{array}{cc}0&v\\
    -1/v&0\end{array}\right)P^{-1}$. These matrices neither commute
nor anticommute with the elements of the commutant of $PN_{\pm
  i}P^{-1}$.

But if $\mu^2\neq 1$, then there exists an
$A=PN_zP^{-1}\in\Lambda^{-1}(\mu^2)$ with $\pm 1\neq z\neq\pm i\}$, so
the commutant $A^c$ of $A$ is an abelian subgroup of
$SL(2,\complex)$, and the anticommutant of $A$ is trivial.
Accordingly, the commutant $\mu^c$ of $\mu$ is the abelian group
$\Lambda(A^c)$.

If $\mu^2=1$, all $z\in\complex$ with $A=PN_zP^{-1}$ and
$\Lambda(A)=\mu$ equal $\pm1$ or $\pm i$. If $z=\pm1$, then $\mu=1$,
and the commutant is $L_1$ and, hence, nonabelian, and if $z=\pm i$,
the above remarks apply.

\smallskip\noindent {\it Proof of (iii).} Since $\mu^2\neq
1\neq\nu^2$, it follows from the preceding statement that the
commutants $\mu^c$ and $\nu^c$ are the maximal-abelian groups
$\{PN_zP^{-1}:\,z\in\dot\complex\}$,
$\{PN_zP^{-1}:\,z\in\dot\complex\}$, or
$\left\{P\left(\begin{array}{cc}1&t\\0&
      1\end{array}\right)P^{-1}:\,t\in\complex\right\}$.  for some
$P\in SL(2,\complex)$.  Consequently, the assumption
$\nu^2\in(\mu^2)^c$ implies $\nu\in(\mu^2)^c$, i.e.,
$\mu^2\in(\nu^2)^c$. This yields the statement by the same argument.
\end{proof}

\begin{lemma}\label{lem:c_and_mu}
Consider $(a,b)\in\M_L$, and shorthand $\lambda(a,b)=:\lambda$.

There exists an element $c\in\hatz$ with $a=\pm j_c b$.
Shorthanding $\lambda(c,b)=:\mu$, one has $\mu^2=\lambda$
and $(a,b)=(\mp\mu b,b)$.
\end{lemma}
\begin{proof} Existence of $c$ follows from Lemma 
\ref{lem:reflecting_planes}. The other statements follow from
\[ j_cj_bj_cj_b=j_{j_cb}j_b=j_aj_b \]
and 
\[ a=\pm j_cb=\mp j_cj_bb=\mp\mu b.\qedhere \]
\end{proof}

\begin{prop_equiv_rel}
$\sim$ is an equivalence relation.
\end{prop_equiv_rel}
\begin{proof} Symmetry and reflexivity are evident, so it remains to prove 
  transitivity. If $\munt\sim\nunt$ and $\nunt\sim\runt$, then
  $\lambda(\munt)=\lambda(\nunt)=\lambda(\runt)=:\lambda$, and there
  exist elements $\mu$ and $\nu$ commuting with $\lambda$ and
  satisfying $\mu^2\munt=\pm\nunt$ and $\nu^2\nunt=\pm\runt$. If
  $\mu^2=1$ or $\nu^2=1$, one trivially has $\munt\sim\runt$. If
  $\nu^2\mu^2=1$, one even has $\munt=\pm\runt$. It follows from
  $\nu^2\mu^2\munt=\pm\runt$ that
\[ \lambda=j_{\nu^2\mu^2 a}j_{\nu^2\mu^2 b}=\nu^2\mu^2j_aj_b\mu^{-2}\nu^{-2}
=\nu^2\mu^2\lambda\mu^{-2}\nu^{-2},\]
and one concludes from Lemma \ref{lem:commutants}.ii that there exists a 
square root $\kappa$ of $\nu^2\mu^2$ commuting with $\lambda$. 
\end{proof}

\subsection{Proof of Proposition \ref{prop:sim_fiber}} \label{sec:sim_fiber}

\begin{lemma}
  $(a,b)\not\sim(a,-b)$ for $(a,b)\in\M_L$, i.e., $g \neq -g$ for all
  $g\in\G_L$.
\end{lemma}

\begin{proof}
  The statement is evident for $b=\pm a$, so it remains to consider
  the case $\lambda(a,b)\neq1$.
  
  Assume $(a,b) \sim (a,-b)$. By Lemma \ref{lem:c_and_mu}, there
  exists an element $\mu \in L_1$ with $\mu^2 = \lambda(a,b)$ and
  $(a,b)=(\pm \mu b,b)$, and by assumption, there exists an element
  $\nu\in L_1$ with $\nu\mu^2\nu^{-1}=\mu^2$ and $\nu^2(a,b)=\pm(a,-b)$.
  $\mu^2$ and $\nu^2$ commute and differ from $1$, so $\mu$ and $\nu$
  commute by Lemma \ref{lem:commutants}.iii.
Assume without loss that $a=\mu b$, then one obtains
\[ (a,b) =(\mu b,b) = \pm\nu^2 (\mu b,-b)
= \pm(\mu \nu^2 b, -\nu^2 b) = \pm(-\mu b,b) = \pm(-a,b), \] leading to the
contradiction $a = -a$ or $b=-b$, respectively.
\end{proof}

\begin{lemma}\label{lem:two-sheeted}  
Suppose that $\lambda(a,b)=\lambda(c,d)$. Then
\begin{itemize}
  \item[(i)] $\lambda(a,c)=\lambda(b,d)$.
  \item[(ii)] $\lambda(a,b)$ and $\lambda(a,c)$ commute.
  \item[(iii)] $(a,b)\sim(c,d)$ or $(a,b)\sim(c,-d)$.
\end{itemize}
\end{lemma}
\begin{proof}[Proof of (i)]
$j_aj_c=j_a(j_cj_d)j_d=j_a(j_aj_b)j_d=j_bj_d$.

\smallskip \noindent {\it Proof of (ii).} 
$j_aj_b\,j_cj_a\,j_bj_a=(j_aj_b)j_c(j_aj_b)j_a=j_cj_dj_cj_cj_dj_a=j_cj_a$.

\smallskip \noindent {\it Proof of (iii).} Since by definition $(a,b)
\sim (-a,-b)$, it suffices to prove $(a,b) \sim (\pm c,\pm d)$ for an
arbitrary choice of signs. If $\lambda(a,b) = 1$ or $\lambda(c,a) = 1$
the statement is trivial. So assume $\lambda(a,b) \neq 1 \neq
\lambda(c,a)$.

By Lemma \ref{lem:c_and_mu}, there exist square roots $\nu_{ab}$ and
$\nu_{cd}$ of $\lambda(a,b)$ and square roots $\nu_{ca}$ and
$\nu_{db}$ of $\lambda(c,a)$ with
\[ a = \pm \nu_{ab}b, \quad c = \pm \nu_{cd} d \quad\text{and}\quad c 
=\pm\nu_{ca}a, \quad d = \pm\nu_{db}b \] for some choice of signs. 

It suffices to prove $\nu_{cd}\nu_{db}=\nu_{db}\nu_{cd}$ and
$\nu_{ab}b = \pm \nu_{cd}b$, since these relations yield the statement
by
\[ (c,d)= (\pm \nu_{cd}\nu_{db} b,\pm\nu_{db} b)= \nu_{db}(\pm\nu_{cd}
b,\pm b) = \nu_{db}(\pm\nu_{ab} b,\pm b) = \nu_{db}(\pm a,\pm b). \]
If $\lambda(a,b)^2\neq 1$, one obtains
$\nu_{cd}\nu_{db}=\nu_{db}\nu_{cd}$ from statement (ii) and Lemma
\ref{lem:commutants} (iii). The remaining condition $\nu_{ab}b = \pm
\nu_{cd}b$ then follows from
\[ j_{\nu_{ab}b}=j_c(j_cj_a)=j_c(j_dj_b)=j_c(\nu_{cd}(j_dj_b)\nu_{cd}^{-1})
=j_c(j_cj_{\nu_{cd}b}))=j_{\nu_{cd}b}. \]
If $\lambda(a,b)^2=1$, one obtains $a=\pm\lambda(a,b)b$ from
$1=j_aj_b\,j_aj_b=j_aj_{-j_ba}$.  Lemma \ref{lem:commutants} (i)
implies $\nu_{ab}^{-1}\nu_{cd} = \lambda(a,b)$ or
$\nu_{ab}^{-1}\nu_{cd} = 1$, proving $\nu_{ab}b = \pm \nu_{cd}b$. The
proof is completed by
\[ \nu_{cd}\lambda(d,b)\nu_{cd}^{-1}=j_cj_{\nu_{cd}b}=j_cj_{\nu_{ab}b}=j_cj_a
=\lambda(d,b) \] and an application of Lemma \ref{lem:commutants} (iv)
yielding $\nu_{cd} \nu_{db} = \nu_{db}\nu_{cd}$ . \qedhere
\end{proof}
One now immediately obtains
\begin{prop_sim_fiber}\label{cor:preciselytwo}
For each $g\in \G_L$ the fiber
  $\tilde\lambda^{-1}(\tilde\lambda(g))$ contains precisely two
  elements.  
\end{prop_sim_fiber}
\begin{proof}
  $g\neq-g$ and $\tilde\lambda(g)=\tilde\lambda(-g)$ for all $g$, so
  each $\tilde\lambda^{-1}(\tilde\lambda(g))$ contains at least two
  elements.
  
  By construction, one has $\lambda(a,b)=\tilde\lambda(g)$ for each
  $(a,b)\in\pi^{-1}(g)$. If $(c,d)\in\M_L$ satisfies
  $\lambda(c,d)=\tilde\lambda(g)=\lambda(a,b)$ as well, Lemma
  \ref{lem:two-sheeted} implies that $(a,b)\sim(c,d)$ or
  $(a,b)\sim(c,-d)$, so $\tilde\lambda^{-1}(\tilde\lambda(g))$
  contains at most two elements.
\end{proof}

\subsection{The sets $Z$ and $\hatz$}

The next goal is the proof of Theorem \ref{thm:G_L}. This requires some
preliminaries. Some properties of the sets $Z$ and $\hatz$ will be
worked out in this section, and the polar decomposition of 
restricted Lorentz transformations into rotations and boosts
will be discussed in the next section.

For each $x\in\reals^{1+3}$, denote the stabilizer of $x$ in $L_1$ as
$\stab(x):=\{\mu\in L_1:\,\mu x=x\}$, and for each subset $M$ of 
$\reals^{1+3}$, define $\stab(M):=\bigcap_{x\in M}\stab(x)$.

\begin{lemma}\label{lem:transitive}
  The actions of $L_1$ on $Z^{+}$ and of $L$ on $Z$ are
  transitive.
\end{lemma}
\begin{proof} Consider any $\xi,\eta\in Z^{+}$. 
  $M_1^+$ is an orbit of $L_1$, so there exists a Lorentz
  transformation $\mu$ with $t_\xi=\mu t_\eta$. This $\mu$ is not
  unique, since $t_\xi=\nu\mu t_\eta$ for each $\nu\in\stab(t_\xi)$.
  
  By construction, one already has $\mu x_\eta\perp t_\xi$, so it
  remains to be shown that $\stab(t_\xi)$ acts transitively on
  $H_1\cap \{t_\xi\}^\perp$. But $\stab(t_\xi)$ is the group of
  rotations with respect to the time vector $t_\xi$, and
  $H_1\cap\{t_\xi\}^\perp$ is the set of time-zero unit vectors, on
  which $\stab(t_\xi)$ acts transitively.

  The second statement now follows from the fact that $-1\in L$.
\end{proof}

\begin{lemma}\label{prop:hat_z_first_countable}
$\hatz$ is a first-countable topological space.
\end{lemma}

\begin{proof}
Let $H$ be a Cauchy surface. Then the set $Z_H:=\{\xi\in
Z^+:\,x_\xi\in H\}$ is a closed subset of $Z^+$.

For each $\xi\in Z^+$, the intersection of the inextendible curve
$\Gamma(\xi)$ with $H$ contains precisely one
element $y_\xi$, and there is a unique generalized boost
$\beta_H(\xi)$ with $y_\xi=\beta_H(\xi) x_\xi$.

Define a map $\zeta_H:\,Z^+\to Z_H$ by $\zeta_H(\xi):=\beta_H(\xi)\xi$.
Then $\xi\,\hatsim\,\eta$ implies $\zeta_H(\xi)=\zeta_H(\eta)$ by
construction, so a map $\hatzeta:\,\hatz\to Z_H$ is well defined by
$\hatzeta(\hatpi(\xi))=\zeta(\xi)$. The diagram
\[\xymatrix{  Z^+ \ar[r]^{\hatpi} \ar[d]_{\zeta_H} 
& \hatz \ar[dl]^{\hatzeta_H} \\ Z_H & }\]
commutes. All maps in this diagram are continuous. This holds for
$\hatpi$ by definition, and it is evident for $\zeta_H$. To show
continuity of $\hatzeta_H$, let $M\subset L_1$ be
open. $\hatzeta_H^{-1}(M)$ is open if and only if
$\pi^{-1}(\hatzeta_H^{-1}(M))$ is open.  This set coincides with
$\zeta_H^{-1}(M)$, which is open by continuity of $\zeta_H$.

Since $\hatzeta_H$ has the continuous inverse $\hatpi|_{Z_H}$, one
finds that $Z_H$ and $\hatz$ are homeomorphic topological spaces.
Since $Z_H$ is first-countable, so is $\hatz$.

\end{proof}

\smallskip\smallskip
One immediately concludes the following corollary.
\begin{corollary}
$\hatz$ and $\M_L$ are Hausdorff spaces.
\end{corollary}

\subsection{Polar decompositions on $L_1$}

The next task will be the proof of Theorem \ref{thm:G_L}, which,
again, is much more involved than its prototype in Ref. \ref{Kuc04}.
A crucial instrument will be the decomposition of Lorentz transformations
into rotations and boosts.  Specify a time direction by distinguishing
a future-directed timelike unit vector $e_0$. 

Consider the euclidean inner product $\langle\cdot,\cdot\rangle_{e_0}$
on $\reals^{1+3}$ defined by 
$\langle x,y\rangle_{e_0}:=-g(x,y)+2g(x,e_0)g(y,e_0)$.
Denote the adjoint of a linear map $T:\,\reals^{1+3}\to\reals^{1+3}$
with respect to this inner product by $T^*$. If $T$ is an
automorphism, then the positive operator
$\betavon(T):=|T|:=(T^*T)^{1/2}$ is a boost, and the orthogonal
operator $\rhovon:=T\cdot|T|^{-1}=T\betavon(T)^{-1}$ is a rotation;
$\betavon(T)$ and $\rhovon(T)$ yield the polar decomposition
$T=\rhovon(T)\betavon(T)$ of $T$.  
On $\G_L$, define $\trho(g):=\rhovon(\tilde\lambda(g))$ and
$\tbeta(g):=\betavon(\tilde\lambda(g))$.

To each time-zero unit vector $e$, assign the class
$\hate:=\hatpi(e_0,e)$. The following lemma immediately follows from
Lemma 2.1 in Ref. \ref{BS}; the proof is recalled here for the
reader's convenience.

\begin{lemma}\label{lem:lambda_onto}
$\tilde\lambda$ is onto.
\end{lemma}
\begin{proof} 
  We prove that $\lambda$ is onto, then the statement follows.
  $\lambda(a,\pm a)=1$ for all $a\in\hatz$, so it remains to show that
  $\lambda^{-1}(\mu)\neq\emptyset$ for each $\mu\neq1$.

Suppose that $\mu=:\rot$ is a rotation, that $\tau$ is a root of
$\rot$, and that $e$ is a time-zero unit vector in the rotation
plane of $\rot$. Then $\rot=\rot j_{\hate}j_{\hate}=j_{\tau\hate}j_{\hate}
=\lambda(\tau\hate,\hate)$.

Suppose that $\mu=:\boost$ is a boost, and let $e$ be a time-zero unit
vector in the fixed-point set of $\boost$. Then
$\boost=j_{\hate}j_{\hate}\beta=j_{\hate}j_{\boost^{-1/2}\hate}=
\lambda(\hate,\boost^{-1/2}\hate)$.

In the remaining case that both $\rhovon(\mu)$ and $\betavon(\mu)$
differ from $1$, the rotation plane of $\rhovon(\mu)$ and the
fixed-point plane of $\betavon(\mu)$ are well-defined two-dimensional
planes contained in the time-zero plane. Since the time-zero plane is
three-dimensional, this implies that the intersection of these planes
is nonempty. Let $e$ be a unit vector in this intersection and let
$\tau$ be a root of $\rhovon(\mu)$. Then
$\mu=\rhovon(\mu)j_{\hate}j_{\hate}\betavon(\mu)
=j_{\tau\hate}j_{\betavon(\mu)^{-1/2}\hate}
=\lambda(\tau\hate,\boost^{-1/2}(\mu)\hate)$.
\end{proof}

\smallskip\smallskip 
Define $\dot R:=R\backslash\{1\}$ and $\dot B:=B\backslash\{1\}$,
and write $\ddot R:=\{\sigma\in R:\,\sigma^2\neq1\}$.

\begin{lemma}\label{lem:equivalent}
  $\rot\in\dot R$ and $\boost\in\dot B$ commute if and only if
  $FP(\rot)=FP(\boost)^\perp$.
\end{lemma}
\begin{proof} Assume $\rot\boost=\boost\rot$.
If $x\in FP(\boost)$, then $\boost\rot
x=\rot\boost x=\rot x$, so $\rot FP(\boost)=FP(\boost)$, 
whence one concludes that either $FP(\boost)=FP(\rot)$
or $FP(\boost)=FP(\rot)^\perp$. Since
$FP(\boost)$ is a spacelike surface, whereas $FP(\rot)$ is timelike,
one concludes $FP(\boost)=FP(\rot)^\perp$. That the condition is 
sufficient, is trivial.
\end{proof}

\begin{lemma}\mbox{}
\label{lem:intersecting_stabilizers} 
\begin{itemize}
\item[(i)] Consider $\mu\in L_1$ with polar decomposition
  $\mu=\rot\boost$.  Then $\rot\boost=\boost\rot$ if and only if there
  exists a time-zero unit vector $e$ with $\mu\in\stab(\hate)$.
 \item[(ii)] Given $a,b\in\hatz$, one has 
$\stab(a)\cap\stab(b)\neq\{1\}$ if and only if $a=\pm b$.
\end{itemize}
\end{lemma}
\begin{proof}[Proof of (i).]
Each rotation or boost is contained in the stabilizer of
$\hate$ for some $e$, so statement (i) trivially holds for
rotation or boosts.

It remains to consider the case that $\rot\neq 1\neq\boost$.
If $\rho\beta=\beta\rho$, then it follows from 
Lemma \ref{lem:equivalent} that the rotation axis of $\rho$
is parallel to the boost direction of $\beta$. Let $e$ be one
of the two unit vectors on this axis, then $\rho$, $\beta$, 
and, hence, also $\rho\beta$ are contained in 
$\stab(W_{\hate})=\stab(\hate)$. So the condition is necessary.

If, conversely, $\mu\in\stab(\hate)$, then there exists a
unique boost $\boostb$ with $\boostb(\mu e_0,\mu e)=(e_0,e)$, and
$\boostb\in\stab(\hate)$ because $\boostb\hate=\boostb\mu\hate=\hate$.
Because $\stab(\hate)$ is abelian, 
$\boostb\mu=\mu\boostb=\rot\boost\boostb$.

The product $\rot\boost\boostb$ has the fixed points $e_0$ and $e$ by
definition of $\boostb$, so it is a rotation, and $\boost\boostb=1$ by
uniqueness of the polar decomposition. As seen above, $\gamma$ commutes
with $\mu$, so $\beta^{-1}$ commutes with $\rho\beta$, i.e.,
$\rho=\beta^{-1}\rho\beta$.

\smallskip {\it Proof of (ii).}  Without loss, assume $a=\hate$.
If $\mu$ is a rotation, then 
$e$ is on the rotation axis of $\mu$, so $FP(\mu)=\hate^{\perp\perp}$,
and the plane $\hate^\perp$ is mapped onto itself. The only other 
time-zero unit vector on the axis of $\mu$ is $-e$, so 
$b=\pm\hate=\pm a$, as stated.

If $\mu$ is a boost, then the
vectors $\ell^{+}:=e+e_0$ and $\ell^{-}:=e-e_0$ are eigenvectors of
$\mu$ associated with distinct eigenvalues $\eps$ and $\eps^{-1}$. The
vectors $\ell^\pm$ are perpendicular to $FP(\mu)$ by invariance of the
metric: if $x\in FP(\mu)$, then
$$\eps\,g(x,\ell^\upparrow) =g(x,\mu\ell^{+})=g(\mu^{-1}x,\ell^{+})
=g(x,\ell^{+}),$$
so $\eps\neq1$ implies $g(x,\ell^{+})=0$, and one
obtains $FP(\mu)=\hate^\perp$.

It remains to consider the case that $\rho\neq 1\neq\beta$. By
Lemma \ref{lem:equivalent}, statement (i) implies $FP(\rot)\perp
FP(\boost)$, so $\ell^\pm$ are fixed points of $\rot$ and, hence
eigenvectors not only of $\beta$, but also of $\mu$.  Additional
eigenvectors in $\hate^\perp$ exist only if $\rho$ is a rotation by
the angle $\pi$; their eigenvalue is $-1$. Since
$\eps\neq-1\neq\eps^{-1}$, the vectors $\ell^\pm$ are the only
eigenvectors of $\mu$ with positive eigenvalues.

By assumption, $\mu\in\stab(b)=:\stab(\hatpi(f_0,f))$, so the polar
decomposition of $\mu$ with respect to $f_0$ commutes. The reasoning
just used yields that the lightlike vectors $f+f_0$ and $f-f_0$ are
eigenvectors of $\mu$ with positive eigenvalues 
and, hence, proportional to $e+e_0$ and $e-e_0$,
respectively, whence $\hate=\pm\hatf$ and, hence, statement (ii) is
obtained.
\end{proof}

\begin{lemma}\label{lem:suppose_that}
  Given any $\mu\in L_1$, suppose that the polar decomposition
  $\mu=\rot_{e_0}\boost_{e_0}$ commutes for all $e_0\in M_1^+$. Then
  $\mu=1$.
\end{lemma}
\begin{proof}
  Because by assumption,
  $\rot_{e_0}\boost_{e_0}=\beta_{e_0}\rot_{e_0}$, there is some
  time-zero unit vector $e$ with $\mu\in\stab(\hate)$.
  
  The subset 
  $$t_{\hate}:=\{d_0\in M_1^+:\,\hatpi(d_0,d)=\hate\quad\mbox{for
    some unit vector $d\perp d_0$}\}$$
  of $M_1^+$ is a hyperbola, so
  there exists some $f_0\in M_1^+\backslash t_{\hate}$.
  
  By assumption, the polar decomposition $\mu=\rot_{f_0}\beta_{f_0}$
  commutes as well, so there is some unit vector $f\perp f_0$ with
  $\mu\in\stab(\hatpi(f_0,f))$. By construction,
  $\hatpi(f_0,f)\neq\pm\hate$, so $W_{\hatpi(f_0,f)}\neq\pm
  W_{\hate}$, whence $\stab(\hatpi(f_0,f))\cap\stab(\hate)=\{1\}$ by
  Lemma \ref{lem:intersecting_stabilizers}.
\end{proof}
For each $(\rot,\boost)\in\dot R\times B$, let $E(\rot,\boost)$ be the
set of all time-zero unit vectors in $FP(\rot)^\perp\cap FP(\boost)$.

\begin{proposition}\label{prop:classification} \mbox{}
\begin{itemize}
  \item[(i)] $E(\rot,\boost)\cong S^1$ if and only if
$\rot\boost=\boost\rot$.
  \item[(ii)] Otherwise, $E(\rot,\boost)=\{\pm e\}$
  for some time-zero unit vector $e$.
\end{itemize}
\end{proposition}
\begin{proof}[Proof of (i)]
  If $\boost=1$, then
  $E(\rot,\boost)=FP(\rot)^\perp\cap\left(\{0\}\times S^2\right)$,
  i.e., the intersection of the time-zero two-sphere with a
  two-dimensional spacelike subspace of the time-zero plane. Such an
  intersection is homeomorphic to $S^1$. If $\rot\neq1\neq\boost$,
  then $\rot\boost=\boost\rot$ if and only if $FP(\boost)\perp
  FP(\rot)$ by Lemma \ref{lem:equivalent}, and this holds if and only
  if $FP(\rot)^\perp\cap FP(\boost)$ is a two-dimensional spacelike
  plane, i.e., if and only if $E(\rot,\boost)$ is homeomorphic with
  $S^1$.

\smallskip
{\it Proof of (ii). }
If $\rot\boost\neq\boost\rot$, then $FP(\rot)^\perp\cap
FP(\boost)$ is not two-dimensional by Lemma
\ref{lem:equivalent}, but since $FP(\rot)^\perp$ and
$FP(\boost)$ are two-dimensional subspaces of the time-zero plane,
their intersection is one-dimensional and contains two opposite
time-zero unit vectors.
\end{proof}

\subsection{Proof of Theorem \ref{thm:G_L}}\label{sect:G_L}

Let $\N^{e_0}$ be the set of all $(\tau,\beta)\in\ddot
R\times B$ with $E(\tau,\beta)\cong\integers_2$ (cf. Prop. 
\ref{prop:classification}). Define a map
$\lambda_1:\,\N^\circe\to L_1$ by
$\lambda_1(\rotb,\boost):=\rotb^2\boost$ and define
$L_1^{e_0}:=\lambda_1(\N^{e_0})$. Furthermore, set 
$\G_L^\circe :=\tilde\lambda^{-1}(L_1^\circe)$. 

For each $\rot\in\ddot R$, there is a unique time-zero unit vector
$\ba(\rot)$ with the property that $\rot$ is a right-handed
rotation with respect to $\ba(\rot)$ by a rotation angle
$\alpha(\rot)$ smaller than $\pi$. The functions $\ba(\cdot)$ and
$\alpha(\cdot)$ are continuous on $\ddot R$, and $\alpha$ has a
continuous extension to a function from all of $R$ onto the closed
interval $[0,\pi]$, we denote this extension by $\alpha$ as well.

For each $\boost\in\dot B$, there exists a unique time-zero unit
vector $\bb(\boost)$ with respect to which $\boost$ is a boost by a
rapidity $\chi(\boost)$ greater than zero. The functions $\bb$ and
$\chi$ are continuous, and the function $\chi$ has a continuous
extension to all of $B$ with values in $\reals^{\geq0}$, which we
denote by $\chi$ as well.

The functions $\tilde\alpha:\,\G_L\to[0,\pi]$ and
$\tilde\chi:\,\G_L\to\reals^{\geq0}$ defined by
$\tilde\alpha(g):=\alpha(\trho(g))$ and
$\tilde\chi(g):=\chi(\tilde\beta(g))$ are continuous.

\begin{lemma} \mbox{}
  \begin{itemize}  
  \item[(i)] The polar decomposition $\rhovon\times\betavon:\,L_1\to
    R\times B$ is continuous.
  \item[(ii)] The restriction of the group product in $L_1$ to
    $R\times B$ is a homeomorphism onto $L_1$.
  \item[(iii)] $N^\circe$ is a two-sheeted covering space of
    $L_1^{\circe}$ when endowed with the covering map $\lambda_1$.
  \end{itemize}
\end{lemma}

\begin{proof}[Proof of (i)] The group product in $L_1$, the map
$\mu\mapsto\mu^*$, and the square-root function are continuous, the
map $\mu\mapsto\betavon(\mu):=\sqrt{\mu^*\mu}$ is continuous. Since the
map $\mu\mapsto\mu^{-1}$ is continuous as well, one concludes that
$\mu\mapsto\rhovon(\mu):=\mu\betavon(\mu)^{-1}$ is continuous. 

\smallskip\noindent
{\it Proof of (ii).} The group product is continuous and inverse to the
continuous polar decomposition. Since the group product is onto, so is
the polar decomposition.

\smallskip\noindent 
{\it Proof of (iii).} $N^\circe$ is an open subset of $\ddot
R\times B$, so it suffices to prove the corresponding statement for
$\ddot R\times B$. So it remains to be shown that $\ddot R$ is a
two-sheeted covering space when endowed with the covering map
$\tau\mapsto\tau^2$. Continuity of this map follows from continuity of
the group product. Conversely, each $\rot\in\ddot R$ has the two
roots $[\ba(\rot),\alpha(\rot/2)]$ and
$[-\ba(\rot),\pi-\alpha(\rot)/2]$, and since $\ba$ and $\alpha$
are continuous maps, the square map has a continuous local inverse.
\end{proof}

\begin{lemma}
  For each $g\in G_L^\circe$, there is a unique square root $\ttau(g)$
  of $\trho(g)$ with $g=\pi(\ttau(g)\hate,\tbeta(g)^{-1/2}\hate)$ for
  both $e\in E(\ttau(g),\tbeta(g))$.
\end{lemma}
\begin{proof}
  If $e\in FP(\tbeta(g))$, then
  $\lambda(\hate,\tbeta(g)^{-1/2}\hate)=\tbeta(g)$.
  If $e\in FP(\trho(g))^\perp$, there are precisely two
  $a\in\hatz$ with $\lambda(a,\hate)=\trho(g)$. Namely, if
  $\tau_\pm$ are the two square roots of the rotation $\trho(g)$,
  then $a_\pm=(\tau_\pm\hate,\hate)$ do the job.
  
  Accordingly, if $e\in E(\trho(g),\tbeta(g))=
  FP(\trho(g))\cap FP(\tbeta(g))^\perp$, the non-equivalent pairs
$\munt^+$ and $\munt^-$ defined by
  $$\munt^\pm:=(\tau_\pm\hate,\hate)\circ(\hate,\tbeta(g)^{-1/2}\hate)=
(\tau_\pm\hate,\tbeta(g)^{-1/2}\hate)$$ satisfy
  $\lambda(\munt^\pm)=\tlambda(g)$. By Corollary
  \ref{cor:preciselytwo}, exactly one of them is contained in
  $\pi^{-1}(g)$.
\end{proof}

Define a ``polar decomposition'' $\eta:\,\G_L^\circe\to\N^\circe$ by
$\eta(g):=(\ttau(g),\tbeta(g))$. Evidently $\eta$ is a bijection, and
the diagram 
\begin{equation}\label{dia:dreieck_circe}
  \xymatrix{ & \G_L^\circe \ar[dl]_{\eta} \ar[d]^{\tilde\lambda}\\
                        \N^\circe \ar[r]_{\lambda_1} & L_1^\circe}
\end{equation}
commutes.
Next define the set
\[ \M_L^\circe:=\{(\tau\hate,\boost^{-1/2}\hate):\,
(\tau,\boost)\in\N^\circe,\,\,e\in E(\tau,\boost)\},\] and define a
map $\lambda_2:\,\M_L^\circe\to\N^\circe$ by
$\lambda_2(\munt):=\eta(\pi(\munt))$. Then the diagrams 
\begin{equation}\label{dia:quadrate_circe}
  \xymatrix{  \M_L^{\circe} \ar[r]^{\pi} \ar[d]_{\lambda_2} & 
\G_L^\circe \ar[dl]_{\eta}     
\ar[d]^{\tilde\lambda} \\
  \N^\circe \ar[r]_{\lambda_1} & L_1^\circe}
  \quad(A)\qquad{\rm and}\qquad
  \xymatrix{  \M_L^{\circe} \ar[r]^{\pi} \ar[d]_{\lambda_2} 
\ar[dr]^{\lambda}& \G_L^\circe    
\ar[d]^{\tilde\lambda} \\
  \N^\circe \ar[r]_{\lambda_1} & L_1^\circe}\quad(B) 
\end{equation}
commute. Define a continuous function 
$\be:R\times B\to H_1\cap e_0^\perp$ by
\[\be(\rot,\boost):=
\frac{\ba(\rot)\times\bb(\boost)}{|\ba(\rot)\times\bb(\boost)|},\]
where $\times$ denotes the vector product within the time-zero plane
$e_0^\perp$.
\begin{lemma} \mbox{}
\begin{itemize} 
    \item[(i)] $\lambda_\circe:=\lambda|_{\G_L^\circe}$ is an open map.

    \item[(ii)] $\lambda_2$ is continuous.

    \item[(iii)] $\eta$ is continuous.
  \end{itemize}
\end{lemma}

\begin{proof}[Proof of (i)] $L_1^\circe$ is first-countable, so it suffices to
  show that for each sequence $(\mu_n)_n$ in $L_1^\circe$ converging
  to a limit $\mu$ and for each $\munt\in\lambda_\circe^{-1}(\mu)$,
  there exists a sequence $(\munt_n)_n$ converging to $\munt$ and
  satisfying $\lambda_\circe(\munt_n)=\mu_n$.

So let $(\mu_n)_n$ be a sequence in $L_1^\circe$ converging to $\mu$. Then
$\rhovon(\mu_n)$ and $\betavon(\mu_n)$ converge to $\rhovon(\mu)$ and
$\betavon(\mu)$, respectively, by continuity of the functions
$\rhovon$ and $\betavon$.  Consequently, the time-zero unit vectors
$e_n:=\be(\rhovon(\mu_n),\betavon(\mu_n))$ tend to the limit
$e=\be(\rhovon(\mu),\betavon(\mu))$. Since $\hatpi$ is continuous, the
sequence $\hate_n$ converges to $\hate$.

Consider, without loss, the element $\munt:=(\tau\hate,
\betavon(\mu)^{-1/2}\hate)$ of the fiber $\lambda^{-1}(\mu)$. There
exists a convergent sequence $(\tau_n)_n$ in $R$ with
$\tau_n^2=\rhovon(\mu_n)$, and the sequence $(\munt_n)_n$ defined by
$\munt_n:=(\tau_n\hate_n,\betavon(\mu_n)^{-1/2}\hate_n)$ satisfies
$\lambda_\circe(\munt_n)=\mu_n$ and $\munt_n\to\munt$.  The same
reasoning applies to the other elements of the fiber
$\lambda_\circe^{-1}(\mu)$.

\smallskip\noindent
{\it Proof of (ii).} For each $\munt_1\in\M_L^\circe$, the fiber
$\lambda_\circe^{-1}(\lambda_\circe(\munt_1))$ contains four elements
$\munt_1,\dots,\munt_4$, and by the Hausdorff property, these have
mutually disjoint open neighborhoods $U_1,\dots, U_4$. Since
$\lambda_\circe$ is open by statement (i), their images are open, so
$V:=\lambda_\circe(U_1)\cap\dots\cap\lambda_\circe(U_4)$ is open.

On the other hand, there is an open neighborhood $Y$ of
$\lambda_2(\munt_1)$ with the property that $\lambda_1|_Y$ is a
homeomorphism onto $W:=\lambda_1(Y)$. Being a covering map,
$\lambda_1$ is open, so $W$ is open.

$V\cap W$ is open, and $\lambda_\circe$ is continuous, so the set
$X:=U_1\cap\lambda_\circe^{-1}(V\cap W)$ is open and contains $\munt_1$.
The diagram 
\[
\xymatrix{ X \ar[d]_{{\lambda_2|}_X} \ar[dr]^{{\lambda_\circe|}_X} & \\ 
Y \ar[r]_-{{\lambda_1|}_Y}& V \cap W}
\]
is a commutative diagram of bijections by construction. Since
$\lambda_\circe|_X$ and $\lambda_1|_Y$ are homeomorphims, so 
is $\lambda_2|_X$.

\smallskip\noindent
{\it Proof of (iii).} Using diagram \ref{dia:quadrate_circe} (B),
one immediately concludes the statement from continuity of $\lambda_2$.
\end{proof}

\begin{lemma}
  $\M_L^\circe$ is a two-sheeted covering space of $N^\circe$ when
  endowed with the covering map $\lambda_2$.
\end{lemma}
\begin{proof} Define continuous maps 
$\munt_\pm:\,N^\circe\to\M_L^\circe$ by
\[ \munt_{\pm}(\tau,\beta):=(\pm\tau\hate(\tau,\beta),
\pm\beta\hate(\tau,\beta)).\] We show that these functions are local
inverses of $\lambda_2$.

For a given $x\in N^\circe$, write $y_\pm:=\munt_\pm(x)$. Since
$\M_L^\circe$ is a Hausdorff space, there exist two disjoint open
neighborhoods $Y_\pm$ of $y_\pm$. By continuity of $\munt_\pm$, the
pre-images $X_\pm:=\munt_\pm^{-1}(Y_\pm)$ are open, and $X:=X_+\cap
X_-$ is an open neighborhood of $x$.  By continuity of $\lambda_2$,
the sets $W_{\pm}:=\lambda_2^{-1}(X)\cap Y_{\pm}$ are open
neighborhoods of $\munt_\pm(x)$ with
$\lambda_2(W_+)=X=\lambda_2(W_-)$. As a consequence, the continuous
maps $\munt_\pm|_X:\,U\to W_{\pm}$ are one-to-one and onto, their
inverse being $\lambda_2$.
\end{proof}

\begin{proposition} \mbox{}
  \begin{itemize}
    \item[(i)] $\eta$ is a homeomorphism.

    \item[(ii)] $\G_L^\circe$ is a Hausdorff space.

    \item[(iii)] $\G_L^\circe$ is a two-sheeted covering space of
$L_1^\circe$ when endowed with the covering map
$\tilde\lambda_\circe$.
  \end{itemize}
\end{proposition}
\begin{proof}  The maps $\pi\circ\munt_+$ and $\pi\circ\munt_-$ coincide
and are inverse to $\eta$ by construction. By continuity of
$\munt_\pm$ and $\pi$, they are continuous. This proves (i) and
implies (ii).

$\tilde\lambda_\circe=\lambda_2\circ\eta$ is a concatenation of
a homeomorphism and a two-sheeted covering map. This yields (iii).
\end{proof}

Next we extend these results to $\dot\G_L$. To this end, recall that
$\mu\in L_1^\circe$ if and only if
$\rhovon(\mu)\betavon(\mu)\neq\betavon(\mu)\rhovon(\mu)$.

\begin{proposition}\mbox{}
  \begin{itemize}
    \item[(i)]  For each $e_0\in M_1^+$, the set $\G_L^\circe$ is an 
open subset of $\dot\G_L$. 
    \item[(ii)] $\bigcup_{e_0\in M_1^+}\G_L^\circe=\dot\G_L$.
    \item[(iii)] $\dot\G_L$ is a two-sheeted covering space of 
$L_1\backslash\{1\}$ when endowed with the covering map
$\tilde\lambda$. 
  \end{itemize}
\end{proposition}

\begin{proof} If a sequence $\mu_n\to\mu$ in $L_1$ with
  $\rhovon(\mu_n)\betavon(\mu_n)=\betavon(\mu_n)\rhovon(\mu_n)$, then
  $\rhovon(\mu)\betavon(\mu)=\betavon(\mu)\rhovon(\mu)$. Namely, one
  has
  $\betavon(\mu_n)^{-1}\rhovon(\mu_n)\betavon(\mu_n)\rhovon(\mu_n)^{-1}=1$
  for all $n$, so
  $\betavon(\mu)^{-1}\rhovon(\mu)\betavon(\mu)\rot^{-1}(\mu)=1$
  follows by continuity of the functions $\beta$, $\rot$,
  $\betavon(\cdot)^{-1}$, and $\rhovon(\cdot)^{-1}$, and of the group
  product.

As a consequence, the set $L_1^\circe$ has a closed complement and,
hence, is an open subset of $L_1$. Accordingly,
$\G_L^\circe=\tilde\lambda^{-1}(L_1^\circe)$ is open by continuity of
$\tilde\lambda$. This proves (i).

It follows from Lemma \ref{lem:suppose_that} that
$\bigcup_{\circe\in M_1^+}L_1^\circe=L_1\backslash\{1\}$, and this
proves statement (ii) by continuity of $\tilde\lambda$.

By statements (i) and (ii), there is, for each $g\in\dot\G_L$, an open
neighborhood restricted to which $\tilde\lambda$ is one-to-one and open.
This proves (iii).
\end{proof}

\begin{proposition} \mbox{} 
  \begin{itemize}
    \item[(i)] $\G_L$ is a Hausdorff space.

    \item[(ii)] $\tilde\lambda$ is open.
  \end{itemize}
\end{proposition}

\begin{proof}[Proof of (i)] Being a union of Hausdorff spaces, $\dot\G_L$ is a
Hausdorff space, so it remains to prove that for each $g$ there are
disjoint neighborhoods $U_1$ and $U_g$ of $1$ and $g\neq1$, respectively,
(which implies that there are disjoint neighborhoods $-U_1$ and $-U_g$ of
$-1$ and $-g$).

$g\neq1$ implies that $(\tilde\alpha(g),\tilde\chi(g))\neq(0,0)$.
Since $\tilde\alpha$ and $\tilde\chi$ are continuous\footnote{with
  respect to the {\it relative} topologies of the {\it closed}
  topological subspaces $[0,\pi]$ and $\reals^{\geq0}$ of $\reals$,}
and since $(\tilde\alpha(h),\tilde\chi(h))=(0,0)$ implies $h=1$,
the open sets
\[\begin{array}{rcccl}
  U_1:=(\tilde\alpha\times\tilde\chi)^{-1}(&[0,\eps)&\times&[0,\eps)&)\\
  {\rm and}\,\,U_g:=(\tilde\alpha\times\tilde\chi)^{-1}(&
  (\tilde\alpha(g)-\eps,\tilde\alpha(g)+\eps)&\times&
  (\tilde\chi(g)-\eps,\tilde\chi(g)+\eps)&)
\end{array}\] 
are disjoint for sufficiently small $\eps>0$.

\smallskip\noindent
{\it Proof of (ii).}  It has been shown that $\dot\G_L$ is a
two-sheeted covering space when endowed with the covering map
$\tilde\lambda$.  Since $\tilde\lambda$ is continuous on all of
$\G_L$, it remains to be shown that $\tilde\lambda$ is open at $\pm1$.
$L_1$ is first countable, so it suffices to show that for each
sequence $\mu_n\to1$ in $L_1$ there exists a sequence $g_n\to1$ in
$\G_L$ with $\tilde\lambda(g_n)=\mu_n$; note that the sequence
$(-g_n)_n$ tends to $-1$ in this case.\footnote{It suffices to
  consider sequences, since $L_1$ is first-countable (which we have
  not yet been proved for $\G_L$ at this stage). Namely, let
  $U_g\subset \G_L$ be a neighborhood of any $g\in\G_L$, and let
  $(\mu_n)_n$ be a sequence in $L_1$ converging to $\tilde\lambda(g)$.
  By assumption there is a sequence $g_n\to G$ with
  $\tilde\lambda(g_n)\to\mu_n$. Since $g_n\to g$ and since $U_g$ is a
  neighborhood of $g$, one has $g_n\in U_g$ for almost all $n$, so
  $\mu_n=\tilde\lambda(g_n)\in\tilde\lambda(U_g)$ for almost all $n$.
  Since this holds for all sequences $\mu_n\to\tilde\lambda(g)$ one
  concludes that $\tilde\lambda(U_g)$ is a neighborhood of
  $\tilde\lambda(g)$ in $L_1$ by first-countability.}  For each $n$
there is a $g_n\in\tilde\lambda^{-1}(\mu_n)$ with
$\tilde\alpha(g_n)\leq\pi/2$. For any $\eps>0$, almost all $g_n$
satisfy $(\tilde\alpha(g_n),\tilde\chi(g_n))\in[0,\pi]\times[0,\eps]$.
Since this is a compact set, the sequence
$(\tilde\alpha(g_n),\tilde\chi(g_n))$ has at least one accumulation
point.  $\tilde\beta(g_n)$ tends to $1$, so $\tilde\chi(g_n)$ tends to
zero, so all accumulation points are in $[0,\pi]\times\{0\}$.

The assumption $\mu_n\to1$ further reduces the set of possible points
to $\{(0,0),(\pi,0)\}$, and opting for $\tilde\alpha(g_n)\leq\pi/2$
rules out $(\pi,0)$. So both $\tilde\alpha(g_n)$ and
$\tilde\chi(g_n)$ tend to zero. It follows that $g_n$ tends to
$1$.
\end{proof}

We now recall and prove the remaining statements made in Sect.
\ref{subsec:G_L_statements}.

\begin{prop_GL_i}
  $\G_L$ is a two-sheeted covering space of $L_1$ when
  endowed with the covering map $\tilde\lambda$.
\end{prop_GL_i}

\begin{proof}
$\dot\G_L$ is a covering of $L_1\backslash\{1\}$ when endowed
with the covering map $\tilde\lambda$, so all that remains to be shown
is that $\tilde\lambda$ is a homeomorphism from some neighborhood $U$
of $1$ or $-1$ onto $\tilde\lambda(U)$.

Since $\G_L$ is a Hausdorff space, there exist disjoint neighborhoods
$U_\pm$ of $\pm1$. Since $\tilde\lambda$ is open, the images
$V_\pm:=\tilde\lambda(U_\pm)$ are open. The intersection $V:=V_+\cap
V_-$ is an open neighborhood of $1\in L_1$, and by continuity of
$\tilde\lambda$, the sets $W_\pm:=U\cap\tilde\lambda^{-1}(V_+\cap
V_-)$ are open and, hence, neighborhoods of $\pm1\in\G_L$,
respectively. Since $W_\pm$ have been constructed in such a fashion
that $\tilde\lambda(W_+)=U=\tilde\lambda(W_-)$, the restrictions
$\tilde\lambda_\pm$ to $W_\pm$ are one-to-one and onto, and since
$\tilde\lambda$ is open, the inverse mappings are continuous. 
\end{proof}

\begin{prop_GL_ii}
$\G_L$ is simply connected.
\end{prop_GL_ii}

\begin{proof} $\hatz$ is pathwise connected, so
$\M_L=\hatz\times\hatz$ is pathwise connected, and since $\pi$ is
continuous, $\G_L=\pi(\M_L)$ is pathwise connected. Since $\G_L$ is a
two-sheeted covering group of $L_1$, and since the fundamental group 
of $L_1$ is isomorphic with $\integers_2$,
one concludes that $\G_L$ is homeomorphic with the universal covering
of $L_1$.
\end{proof}

\begin{prop_GL_iii}
There is a unique group product $\odot $ on ${\G}_{L}$ with
  the property that the diagram
\begin{equation}\label{dia:thm_G_L}
  \xymatrix{ {\M}_{L}\times {\M}_{L}  \ar[r]^-{\circ } \ar[d]_{\pi\times\pi} 
        &  {\M}_{L}\ar[d]^{\pi} \\
  {\G}_{L}\times {\G}_{L} \ar[r]^-{\odot} 
\ar[d]_{\tilde\lambda\times\tilde\lambda} 
        & {\G}_{L} \ar[d]^{\tilde\lambda} \\
  {L}_1\times {L}_1 \ar[r]^-{\cdot}            &{L}_1 }
\end{equation}
commutes.
\end{prop_GL_iii}
\begin{proof}
  The outer arrows of the diagram commute, so it suffices to prove
  existence and uniqueness of a group product conforming with the
  lower part. But it is well known that each simply connected covering
  space $\tilde G$ of a topological group $G$ can be endowed with a
  unique group product $\odot$ such that $G$ is a covering
  group.\footnote{See, e.g., Props. 5 and 6 in Sect. I.VIII. in Ref.
    \ref{Chev}.}
\end{proof}
\begin{lem_adjoint}
Given $h\in\G_L$ and $(c,d)\in\M_L$, one has 
\begin{equation}
h\pi(c,d)h^{-1}=\pi\(\tlambda(h)c,\tlambda(h)d\).
\end{equation}
\end{lem_adjoint}
\begin{proof}
The function $F:\,G_L\to\G_L$ defined by 
\[F(h):=\pi\(\tlambda(h)c,\tlambda(h)d\)^{-1}\,h\pi(c,d)h^{-1}\]
has the property that $\lambda(F(h))=1$ and that, hence, it
takes values in the discrete set $\{\pm1\}\subset\G_L$. Since $F$ is
continuous and $L_1$ is connected, $F$ is constant, and because
$F(1)=1$, it follows that $F(h)=1$ for all $h$.
\end{proof}

\section{Spin \& Statistics}\label{sect:spst}

The preceding section has provided the basis of a general
spin-statistics theorem, which is the subject of this section. From an
intrinsic form of symmetry under a charge conjugation combined with a
time inversion and the reflection in {\it one} spatial direction,
which is referred to as {\it modular P$_1$CT-symmetry}, a strongly
continuous unitary representation $\tilde W$ of $\G_L$ will be
constructed. It is, then,
elementary to show that $\tilde W$ exhibits Pauli's spin-statistics
relation.

\smallskip\smallskip 
Let $F$ be an arbitrary quantum field on $\reals^{1+3}$ in a Hilbert
space $\H$. The standard properties of relativistic quantum
field to be used here are practically the same as in Ref. \ref{Kuc04}
and are recalled here for the reader's convenience.
\begin{itemize}
\item[(A)] {\it Algebra of field operators.} Let $\C$ be a linear
  space of arbitrary dimension,\footnote{Again, $\C$ is the
    ``component space'', and its dimension equals the number of
    components, which may be infinite in what follows.} 
and denote by
  $\test$ the space $C_0^\infty(\reals^{1+3})$ of test functions on
  $\reals^{1+3}$. The field $F$ is a linear function that assigns to
  each $\fphi\in\C\otimes\test$ a linear operator $F(\fphi)$ in a
  separable Hilbert space $\H$. 
\begin{itemize}
\item[(A.1)] $F$ is free from redundancies in $\C$, i.e., if
  ${\mathfrak c},\md\in\C$ and if $F(\mc\otimes\phi)=F(\md\otimes\phi)$ for
  all $\phi\in\test$, then $\mc=\md$.
\item[(A.2)]  Each field operator $F(\fphi)$ and its
  adjoint $F(\fphi)^\dagger$ are densely defined. There exists a dense
  subspace $\D$ of $\H$ contained in the domains of $F(\fphi)$ and
  $F(\fphi)^\dagger$ and satisfying $F(\fphi)\D\subset\D$ and
  $F(\fphi)^\dagger\D\subset\D$ for all $\fphi\in\C\otimes\test$.
\end{itemize}  
Denote by ${\bf F}$ the algebra generated by all $F(\fphi)|_\D$ and
  all $F(\fphi)^\dagger|_\D$.  Defining an involution $*$ on ${\bf F}$
  by $A^*:=A^\dagger|_\D$, the algebra ${\bf F}$ is endowed with the
  structure of a $*$-algebra.
  
  Let $\bF({a})$ be the algebra generated by all
  $F(\mc\otimes\varphi)|_\D$ and all
  $F(\mc\otimes\varphi)^\dagger|_\D$ with $\supp(\varphi)\subset W_a$,
  where $W_a$ denotes, as above, the Rindler wedge of $a$.  The
  algebra $\bF(a)$ inherits the structure of a $*$-algebra from $\bF$
  by restriction of $*$.
\begin{itemize}
\item[(A.3)] $\bF(a)$ is
nonabelian for each $a$, and $a\neq b$ implies $\bF(a)\neq\bF(b)$.
\end{itemize}

\item[(B)] {\it Cyclic vacuum vector.} There exists a vector
  $\Omega\in\D$ that is cyclic with respect to each $\bF(a)$.

\item[(C)] {\it Normal commutation relations.} There exists a unitary
   and self-adjoint operator ${k}$ on $\H$ with $k\Omega=\Omega$ and
   with $k\bF({a})k=\bF({a})$ for all $a$. Define
   $F_\pm:=\frac{1}{2}(F\pm{k} F{k})$.  If $\mc$ and $\md$ are
   arbitrary elements of $\C$ and if
   $\varphi,\psi\in\test$ have spacelike separated
   supports, then
\begin{align*}
  F_+(\mc\otimes\phi)F_+(\md\otimes\psi)
&=F_+(\md\otimes\psi)F_+(\mc\otimes\phi),\\
  F_+(\mc\otimes\phi)F_-(\md\otimes\psi)
  &=F_-(\md\otimes\psi)F_+(\mc\otimes\phi),\quad{\rm and}\\
  F_-(\mc\otimes\phi)F_-(\md\otimes\psi)
&=-F_-(\md\otimes\psi)F_-(\mc\otimes\phi)
\end{align*}
for all $\mc,\md\in\C$.
\end{itemize}
The involution ${k}$ is the {\it statistics operator}, and $F_\pm$ are
the bosonic and fermionic components of $F$, respectively. Defining
$\kappa:=(1+ik)/(1+i)$ and $F^t(\md\otimes\psi):=\kappa
F(\md\otimes\psi)\kappa^\dagger$, the normal commutation relations
read
\[\left[F(\mc\otimes{\varphi}),F^t(\md\otimes\psi)\right]=0.\]
This property is referred
to as {\it twisted locality}. Denote
$\bF(a)^t:=\kappa\bF(a)\kappa^\dagger$.

\smallskip\smallskip These properties imply that $\Omega$ is {\it
  separating} with respect to each algebra $\bF({a})$, i.e., there is
no nonzero operator $A\in\bF({a})$ with $A\Omega=0$.\footnote{For
  details on this and the following statements, see Ref. \ref{Kuc04}
  or \ref{BR1}.} As a consequence, an antilinear operator
$R_{a}:\,\bF({a})\Omega\to\bF({a})\Omega$ is defined by
$R_{a}A\Omega:=A^*\Omega$. This operator is closable.  Its closed
extension $S_{a}$ has a unique polar decomposition
$S_{a}=J_{a}\Delta_{a}^{1/2}$ into an antiunitary operator $J_{a}$,
which is called the {\it modular conjugation}, and a positive operator
$\Delta_{a}^{1/2}$, which is called the {\it modular operator}. $J_a$
is an involution.\footnote{ $S_{a}$, $J_{a}$, and $\Delta^{1/2}_{a}$
  are the objects of the well-known {\it modular theory} developed by
  Tomita and Takesaki.}

\smallskip\smallskip For each $a\in\hatz$, let $j_a$ be the
orthogonal reflection at the edge of $W_a$.

\begin{itemize}
\item[(D)]{\it Modular P$_1$CT-symmetry.} For each $a\in\hatz$, there
  exists a linear involution $C_a$ in $\C$ such that for all
  $\mc\in\C$ and $\varphi\in\test$, one has
  \[J_{a}F(\mc\otimes\varphi)J_{a}=F^t(C_a\mc\otimes
  \overline{j_a\varphi}).\]
  The map $a\mapsto J_a$ is strongly
  continuous.\footnote{If one assumes covariance with respect to some
    strongly continuous representation of $\G_L$ (which may also
    violate the spin-statistics connection), this is straightforward
    to derive. But covariance, as such, is not needed.}
\end{itemize}
It will now be recalled that pairs of modular P$_1$CT-reflections give rise to
a strongly continuous representation of $\G_L$ which exhibits Pauli's
spin-statistics connection.

\begin{lemma}\label{lem:com_rel}
  Let $K$ be a unitary or antiunitary operator in $\H$ with 
  $K\Omega=\Omega$, and suppose there are
  $a,b\in \hatz $ such that $K\bF({a})K^\dagger=\bF({b})$. Then
  $KJ_{a}K^\dagger=J_{b}$, and $K\Delta_a K^\dagger=\Delta_b$.
\end{lemma}
\begin{proof} If $B\in\bF(b)$, then $KS_{a}K^\dagger
  B\Omega=KS_{a}\underbrace{K^\dagger BK}_{\in\bF({a})}\Omega
  =B^*\Omega=S_{b}B\Omega$. The statement now follows by uniqueness of
  the polar decomposition.
\end{proof}

This lemma yields a couple of important relations. For each $a\in\hatz$,
one has $k\bF(a)k^\dagger=k\bF(a)k=\bF(a)$, so 
\begin{equation}\label{kJk}
kJ_ak=J_a,\quad{\rm whence}\quad J_a\kappa=\kappa^\dagger J_a
\end{equation}
follows by antilinearity of $J_a$.
By modular P$_1$CT-symmetry, 
$J_a\bF(a)J_a=\bF^t(-a)=\kappa\bF(-a)\kappa^\dagger$, so 
\begin{equation}\label{minus a}
J_a=J_aJ_aJ_a=\kappa J_{-a}\kappa^\dagger.
\end{equation}
It also follows from modular P$_1$CT-symmetry that 
$J_a\bF(b)J_a=\bF^t(j_ab)=\bF(-j_ab)$, so 
\begin{equation}\label{eq:reflections}
J_{a}J_{b}J_{a}=J_{-j_{a}{b}}=J_{j_aj_bb}=J_{\lambda(a,b)b}.
\end{equation}
These consequences of Lemma \ref{lem:com_rel} will be used extensively
in what follows without further mentioning.

For every $(a,b)\in\M_L$, define $W(a,b):=J_aJ_b$.

\begin{theorem} \mbox{} 
  \begin{itemize}
    \item[(i)] There exists a representation $\tilde W:\,\G_L\to W(\M_L)$ in
  $\H$ with the property that the diagrams
\begin{equation}\label{diagram W}
  \begin{xy}
    \xymatrix{ {\M}_{L}\times {\M}_{L} \ar[r]^-{\circ}
      \ar[d]_{\pi\times\pi} &  {\M}_{L}\ar[d]^{\pi} \\
      \G_L\times\G_L \ar[r]^-{\odot} \ar[d]_{\tilde W\times\tilde W} &
      \G_L \ar[d]^{\tilde W} \\
      \tilde W(\G_L)\times\tilde W(\G_L) \ar[r]^-{\cdot}
      \ar[d]_{\lambda_W\times\lambda_W}
      & \tilde W(\G_L)  \ar[d]^{\lambda_W} \\
      L_1\times L_1 \ar[r]^-{\cdot} & L_1 } \end{xy} \qquad{\rm
    and}\qquad
\begin{xy}
\xymatrix{  {\M}_{L} \ar[r]^{\pi}\ar[d]_-{W} & {\G}_{L} \ar[dl]^-{\tilde W} \\
    W(\M_L) & }
\end{xy}
\end{equation}
commute.

\item[(ii)] There is a representation $\tilde{D}$ of $\G_L$ in $\C$
  such that
\begin{equation}\label{modular covariance}
\tilde W(g)F(\mc\otimes\phi)\tilde W(g)^*=
F(\tilde{D}(g)\mc\otimes\tilde\lambda(g)\phi)\quad 
\forall g,\mc,\phi,
\end{equation}
where $\tilde\lambda(g)\phi:=\phi(\tilde\lambda(g)^{-1}\,\cdot)$.
  \end{itemize}
\end{theorem}
\begin{proof}[Proof of (i)]
Fix some $e_0\in M_1^+$. 

For each $r\in\G_L$ with $\tilde\lambda(r)\in R$, there exists a
unique rotation $\tau$ with $\tau^2=\tilde\lambda(r)$ and
$r=\pi(\tau\hate,\hate)$ for all time-zero unit vectors in
$FP(\lambda(r))^\perp$. For each $\nunt:=(\hate',\hatf')\sim\munt$,
there exists a rotation $\rot$ with
$\rot\lambda(\munt)\rot^{-1}=\lambda(\munt)$ and $\rot^2\munt=\nunt$.
Because $a\mapsto J_a$ and, hence, also the map $W$ is continuous by
assumption of modular P$_1$CT-symmetry, one can mimick the proof of
Lemma 2.4 in Ref. \ref{BS} in order to show that $W(\munt)=W(\nunt)$,
and one can define a unitary operator $\tilde W_{e_0}(r)$ by $\tilde
W_{e_0}(r):=W(\munt)$. It has been shown in Ref. \ref{Kuc04} that
these operators give rise to a representation of the subgroup
$\G_R:=\tilde\lambda^{-1}(R)$ of $\G_L$.

For each $b\in\G_L$ with $\tlambda(b)\in B$, the class $\pi^{-1}(b)$
contains elements of the form either $(\hate,\boost\hate)$ or
$(\hate,-\boost\hate)$, where $\boost:=\tilde\beta(b)^{-1/2}$. The
one-parameter group of rotations around the boost direction of
$\boost$ acts transitively on the set of such elements. If $\nunt$ is
a second such argument equivalent to $\munt$, one can, again, use the
reasoning of Ref.  \ref{BS} in order to show that $W(\munt)=W(\nunt)$,
and one can define $\tilde W_{e_0}(b):=W(\munt)$. Furthermore, if
$(b_t)_t$ is a one-parameter subgroup of $\G_L$ with $\tlambda(b_t)\in
B$, it follows from the results of Ref. \ref{BS} that $\tilde
W(b_s)\tilde W(b_t)=\tilde W(b_{s+t})$.

The polar decomposition in $L_1$ can be lifted to a polar
decomposition in $\G_L$. Namely, given an arbitrary $g\in\G_L$, there
exist $r_g,b_g\in\G_L$ with $\tlambda(r_g)=\rhovon(\tlambda(g))$ and
$\tlambda(b_g)=\betavon(\tlambda(g))$ and $r_gb_g=g$. This
decomposition is unique up to replacement of $r_g$ and $b_g$ by $-r_g$
and $-b_g$, respectively.

Therefore, the operator $\tilde W(g):=\tilde W(r_g)\tilde W(b_g)$ does
not depend on the choice of this polar decomposition.

For arbitrary $g=r_gb_g\in\G_L$, define $\tilde W_{e_0}(g):=\tilde
W_{e_0}(r_g)\tilde W_{e_0}(b_g)$.  Note that the definition of $\tilde
W_{e_0}(g)$ depends on $e_0$ as it stands; but the index will be
dropped for the time being.  

\begin{lemma}\label{lem:comm_rot_boost}
  Consider $g,h\in\G_L$ with $\tlambda(g),\tlambda(hgh^{-1})\in B$
and $\tlambda(h)\in R\cup B$. Then
$$\tilde W(h)\tilde W(g)\tilde W(h)^*=\tilde W\(hgh^{-1}\).$$
\end{lemma}
\begin{proof}
  It follows from eq.  (\ref{eq:reflections}) and Lemma
  \ref{lem:adjoint_action} that for each $(a,b)\in\pi^{-1}(g)$, one
  has
\begin{equation}\label{eq:adjoint_W}
\begin{split}
  \tilde W(h)\tilde W(\pi(a,b))\tilde W(h)^*&= \tilde W(h)J_aJ_b\tilde
  W(h)^*= J_{\tlambda(h)a}J_{\tlambda(h)b}\\&
  =W(\tlambda(h)a,\tlambda(h)b)=\tilde W\(\pi\(\tlambda(h)a,\tlambda(h)b\)\)\\
  &=\tilde W(h\pi(a,b)h^{-1}).
\end{split}
\end{equation}
\end{proof}

\begin{lemma}\label{lem:stabilizer_W(munt)}
If $\tilde\lambda(g)\in\stab(a)$ for some $a\in\hatz$,
then $W(\munt)=\tilde W(g)$ for all $\munt\in\pi^{-1}(g)$.
\end{lemma}
\begin{proof}
Without loss, suppose that $a=\hate$ for some time-zero unit vector $e$.
If $g=r_gb_g$ and $h=r_hb_h$ with
$\tilde\lambda(g),\tilde\lambda(h)\in\stab(\hate)$ for some time-zero
unit vector $e$, then Lemma \ref{lem:comm_rot_boost} implies 
\begin{equation}\label{eq:hgh}
\begin{split}
\tilde W(h)\tilde W(g)\tilde W(h)^*&=\tilde W(r_h)\tilde
W(b_h)\cdot\tilde W(r_g)\tilde W(b_g)\cdot\tilde W(b_h)^*\tilde W(r_h)^*\\
&=\tilde W(g)
\end{split}
\end{equation}
Let $\munt\in\M_L$ satisfy $W(\munt)=\tilde W(g)$.  If
$\nunt\sim\munt$ and $\nunt \neq\munt$, then there exists, by
definition of $\sim$, a $\mu\in L_1$ with $\mu^2\neq 1$, commuting
with $\tilde\lambda(g)$ and satisfying $\mu^2\munt=\pm \nunt$. Since
$\stab(\hate)$ is a maximal abelian group and since
$\tilde\lambda(g)\in\stab(\hate)$ by assumption, one concludes
$\mu\in\stab(\hate)$, and for each $h$ with $\tilde\lambda(h)=\mu$,
one obtains from eq. (\ref{eq:hgh})
\[W(\nunt)=W(\pm\mu^2\munt)=W(\mu^2\munt)
=\tilde W(h)\tilde W(g)\tilde W(h)^*=\tilde
W(g) =W(\munt).\]
\end{proof}

\smallskip\smallskip\noindent 
{\it Proof of (i) (contd.).} Next let $g\in\G_L$
be arbitrary with polar decomposition $r_gb_g$.

$\tilde W(g)$ is an element of $W(\M_L)$. Namely, recall that there
exist a time-zero unit vector $e$ and a rotation $\tau$ such that
$g=\pi(\tau\hate,\beta^{-1/2}\hate)$, where
$\beta=\tilde\lambda(b_g)\in B$. One concludes
\begin{align*}
  \tilde W(g)&=\tilde W(r)\tilde W(b)
=J_{\tau\hate}J_{\hate}J_{\hate}J_{\beta^{-1/2}\hate}\\
  &=J_{\tau\hate}J_{\beta^{-1/2}\hate}\in W(\M_L)
\end{align*}
$\tilde W$ is a representation. Namely,
\begin{align*}
  \tilde W(g)\tilde W(h)&=\tilde W(r_g)\tilde W(b_g)\tilde
  W(r_h)\tilde W(b_h)\\&=\tilde W(r_g)\tilde W(r_h)\,\left(\tilde
  W(r_h)^*\tilde W(b_g)\tilde W(r_h)\right)\,\tilde W(b_h)\\
&=:\tilde W(r_gr_h)\,\tilde W(b_f)\tilde W(b_h)
\end{align*}
The last two terms implement the generalized boost
\[\tilde\lambda(b_f)\tilde\lambda(b_h)
=\tilde\lambda(b_f)^{1/2}\left(\tilde\lambda(b_f)^{1/2}
\tilde\lambda(b_g)\tilde\lambda(b_f)^{1/2}\right)
\tilde\lambda(b_f)^{-1/2},\]
so Lemma \ref{lem:stabilizer_W(munt)} yields 
\begin{align*}
\tilde W(b_fb_h)&=J_{\pm\tilde\lambda(b_f)^{1/2}\hate}
J_{\tilde\lambda(b_h)^{-1/2}\hate}\\&=
J_{\pm\tilde\lambda(b_f)^{1/2}\hate}J_{\hate}^2
J_{\tilde\lambda(b_h)^{-1/2}\hate}=
J_{\hate}J_{\pm\tilde\lambda(b_f)^{-1/2}\hate}
J_{\hate}J_{\tilde\lambda(b_h)^{-1/2}\hate}\\&=
\tilde W(b_f)\tilde W(b_h).
\end{align*}
Now write $b_fb_h=:d=r_db_d$, then
\begin{align*}
\tilde W(g)\tilde W(h)&=\tilde W(r_gr_hr_d)\tilde W(b_d)\\
&=\tilde W(r_gr_hr_db_d)\\
&=\tilde W(gh).
\end{align*}

\smallskip\noindent
{\it Proof of (ii).} Define a map $D$ from $\M_L$ into the
automorphism group $\Aut(\C)$ of $\C$ by $D(a,b):=C_aC_b$. If
$(a,b)\sim(c,d)$, then modular P$_1$CT-symmetry implies
\begin{align*}
F(C_aC_b\mc\otimes j_aj_b\phi)&=W(a,b)F(\mc\otimes\phi)W(a,b)^*\\
&=W(c,d)F(\mc\otimes\phi)W(c,d)^*\\&
=F(C_cC_d\mc\otimes j_cj_d\phi)\\&
=F(C_cC_d\mc\otimes j_aj_b\phi)
\end{align*}
for all $\mc$ and all $\phi$. Using assumption (A.1), one obtains
$C_aC_b\mc=C_cC_d\mc$ for all $\mc$, so $D(a,b)=D(c,d)$, and a map
$\tilde D:\,\G_L\to\Aut(\C)$ is defined by $\tilde
D(\pi(\munt)):=D(\munt)$. This map $\tilde D$ now inherits the
representation property from $\tilde W$. 
\end{proof}

\begin{theorem}[Spin-statistics connection]
\[F_\pm(\mc\otimes\phi)=\frac{1}{2}(1\pm F(\tilde {D}({{-1}})\mc\otimes\phi)\]
for all $\mc$ and all $\phi$.
\end{theorem}
\begin{proof}
For each $a\in \hatz$ one has
\[\tilde W({{-1}})=J_aJ_{-a}=J_a\kappa
J_a\kappa^\dagger=J_a^2(\kappa^\dagger)^2 =k,\]
so
\begin{align*}
kF(\mc\otimes\varphi)k&
=\tilde W({{-1}})F(\mc\otimes\varphi)\tilde W({{-1}})\\&
=\tilde
W({{-1}})F(\mc\otimes\varphi)\tilde W({{-1}})^\dagger =\tilde
F({\tilde D}({{-1}})\mc\otimes\varphi).\qedhere
\end{align*}
\end{proof}

\bigskip If, in particular, $\tilde{D}$ is irreducible with spin $s$,
then $\tilde{D}(-1)=e^{2\pi i s}$, so $F_-=0$ for integer $s$ and
$F_+=0$ for half-integer $s$.

\section{Other modular symmetries}\label{sect:other}

Evidently, the operator $\Theta:=J_{\hate_1}J_{\hate_2}J_{\hate_3}$
implements a full PCT-symmetry. $\Theta$ depends on the handedness of
the triple $(e_1,e_2,e_3)$ only \cite{Kuc04}.

\smallskip\smallskip

As mentioned earlier, Guido and Longo obtained a spin-statistics
theorem in the above spirit in Ref. \ref{GL95}. Instead of the
P$_1$CT-reflections, they assumed the modular groups associated with
the algebras $\bF(a)$ and the vacuum vector, which satisfy the
KMS-condition, to implement Lorentz boosts --- which is the abstract
verson of the Unruh effect. This suffices to construct a
representation of $L_1$ not from P$_1$CT-reflections, but from the
one-parameter groups implementing the boosts (for which the
commutation relations requested for covariance are not assumed from
the outset \cite{BGL93,GL95}). This representation can, then easily be
shown to satisfy Pauli's spin-statistics relation.

Since both the above and their representation have been constructed from
the basic elements of Tomita-Takesaki theory, one should expect them to
coincide. Indeed, this is he case.

Denote by $\Lambda_a$ the unique one-parameter group of Lorentz boosts
that map the wedge $W_a$ onto itself for each $a\in\hatz$.  $F$
exhibits the Unruh effect if and only if for each $a\in\hatz$ there
exists a one-parameter group $V'$ of internal symmetries of $F$ with
$\Delta_a^{it}F(x)\Delta_a^{-it}= V'(\Lambda_a(t))F(\Lambda_a(-t)x)$.
In this case the modular unitaries $\Delta_a^{it}$, $a\in\hat{Z}$,
$t\in\reals$, generate a covariant unitary representation $U'$ of
$\widetilde L_1$ \cite{BGL93}.  This representation exhibits Pauli's
spin-statistics connection \cite{GL95}.

On the other hand, the Unruh effect implies modular P$_1$CT-symmetry
\cite{GL95}, and, hence, yields the representation $W$ constructed above
as well.

\begin{lemma}
$U'=W$.
\end{lemma}
\begin{proof}  It suffices to show that for each $\Delta_a^{it}$,
$a\in\hat{Z}$, $t\in\reals$, there exist $b,c\in\hat{Z}$ such that
$\Delta_a^{it}=J_bJ_c$.

If for some $a,b\in\hat{Z}$ there exist zweibeine
$\xi\in\hatpi^{-1}(a)$ and $\eta\in\hatpi^{-1}(b)$ with $t_\xi=t_\eta$
and $x_\xi\perp x_\eta$, then it follows from Lemma \ref{lem:com_rel}
that $J_a\Delta_a^{it}J_a=\Delta_a^{-it}$ and
$\Delta_a^{-it/2}J_a\Delta_a^{it/2}=J_{\Lambda_a(t/2)}$ for all
$t\in\reals$, so
\[J_a\Delta_a^{it}=\Delta_a^{-it/2}J_a\Delta^{it/2}=J_{\Lambda_a(-t/2)a},
\quad\mbox{i.e.,}\quad \Delta_a^{it}=J_aJ_{\Lambda_a(-t/2)a}. \qedhere
\]
\end{proof}

\section*{Conclusion}

Both the classical geometry and the fundamental quantum field
theoretic representations of the restricted Lorentz group ${L}_1$ are
based on reflection symmetries. At the classical level, a simply
connected covering group $\G_L$ of ${L}_1$ can be constructed from
P$_1$T-reflections. 

For a typical quantum field $F$, a class of antiunitary
P$_1$CT-operators exists that are fixed by the intrinsic structure of
the respective field. These are the fundamental symmetries of quantum
field theories, and they give rise to a unitary representation of the
Lorentz group. In order to show this, the existence of such a
representation does not need to be assumed from the outset. On the
other hand, the construction yields a distinguished representation of
the Lorentz group even in cases where several covariant
representations are present.

It may happen in such cases that representations satisfying Pauli's
relation coexist with representations violating it. In any case, the
representation constructed from the modular P$_1$CT-conjugations
exhibits the right spin-statistics connection, and this is,
eventually, straightforward to see.

\subsection*{Acknowledgements}
Thanks are due to Professor D. Arlt for critically reading preliminary
versions of the manuscript.

The authors have been supported by the Emmy-Noether-programme of the
Deutsche Forschungs\-gemeinschaft. R. L. also acknowledges support
from the Graduiertenkolleg ``Zuk\"unftige Entwicklungen der
Elementarteilchenphysik'' at the University of Hamburg.

\end{document}